\documentclass[aps, prb, reprint, superscriptaddress]{revtex4-2}

\pdfoutput=1 
\usepackage{graphicx}
\usepackage{color,amsmath}
\DeclareMathOperator{\arcosh}{arcosh}
\usepackage[normalem]{ulem}
\usepackage[utf8]{inputenc}

\usepackage[colorlinks,linkcolor=blue,citecolor=blue,urlcolor=blue]{hyperref}
\usepackage[separate-uncertainty = true]{siunitx}
\usepackage{comment}
\usepackage{layouts}
\usepackage[acronym]{glossaries}
\glsdisablehyper

\begin{document}
\title{Characterization of quasiparticle tunneling in a quantum dot from temperature dependent transport in the integer and fractional quantum Hall regime}
% can we adress the physics more in the title: Characterization of tunneling processes through quantum dots in the in and fq Hall regime from temperature dependent transport

\author{Marc~P.~Röösli}
\email{marcro@phys.ethz.ch}
\affiliation{Solid State Physics Laboratory, Department of Physics, ETH Zurich, 8093 Zurich, Switzerland}

\author{Michael~Hug}
\affiliation{Solid State Physics Laboratory, Department of Physics, ETH Zurich, 8093 Zurich, Switzerland}

\author{Giorgio~Nicolí}
\affiliation{Solid State Physics Laboratory, Department of Physics, ETH Zurich, 8093 Zurich, Switzerland}

\author{Peter~Märki}
\affiliation{Solid State Physics Laboratory, Department of Physics, ETH Zurich, 8093 Zurich, Switzerland}

\author{Christian~Reichl}
\affiliation{Solid State Physics Laboratory, Department of Physics, ETH Zurich, 8093 Zurich, Switzerland}

\author{Bernd~Rosenow}
\affiliation{Institute for Theoretical Physics, Leipzig University, Leipzig D-04009, Germany}

\author{Werner~Wegscheider}
\affiliation{Solid State Physics Laboratory, Department of Physics, ETH Zurich, 8093 Zurich, Switzerland}

\author{Thomas~Ihn}
\affiliation{Solid State Physics Laboratory, Department of Physics, ETH Zurich, 8093 Zurich, Switzerland}

\author{Klaus~Ensslin}
\affiliation{Solid State Physics Laboratory, Department of Physics, ETH Zurich, 8093 Zurich, Switzerland}

\date{\today}

% Glossaries for abbreviations
\newacronym[longplural={quantum dots}]{QD}{QD}{quantum dot}
\newacronym[longplural={quantum point contacts}]{QPC}{QPC}{quantum point contact}
\newacronym[longplural={charge stability diagrams}]{CSD}{CSD}{charge stability diagram}
\newacronym[longplural={charge detectors}]{CD}{CD}{charge detector}
\newacronym[longplural={two-dimensional electron gases}]{2DEG}{2DEG}{two dimensional electron gas}

\begin{abstract}
We report on magnetoconductance measurements through a weakly coupled quantum dot, containing roughly 900 electrons, in a wide magnetic field range from 0~T to 12~T. We find modulations of the conductance resonances in the quantum Hall regime for higher integer filling factors $6> \nu_\mathrm{dot} > 2$, in addition to modulations at $2> \nu_\mathrm{dot} > 1$ and at fractional filling factors $\nu_\mathrm{dot} \gtrsim 2/3$, $1/3$. Depending on the internal filling factor, edge reconstruction inside the quantum dot leads to the formation of multiple concentric compressible regions, which contain discrete charge and are separated by incompressible rings. Quasiparticle tunneling between different compressible regions results in magnetic-field-(pseudo)-periodic modulations of the Coulomb resonances with different periodicities, additional super-periodicity or non-periodic features. The evolution of the period in magnetic field indicates cyclic depopulation of the inner compressible regions close to integer filing factors. We study the temperature dependence of the conductance resonances for different magnetic fields. While at low fields, the resonance amplitude decays inversely with temperature, corresponding to single-level transport, the temperature dependence of the amplitude evolves continuously into a linearly increasing behavior for dot filling factor $2> \nu_\mathrm{dot} > 1$. This coincides with a reduction of the charging energy by approximately a factor of 2 in the same regime. At zero magnetic field, the temperature dependence differs for individual resonances, and is found to depend on the closeness to quantum dot states that couple strongly to the leads. 
The presented experiments complement and extend previous results on internal rearrangements of quasiparticles for weakly coupled quantum dots in magnetic field.

\end{abstract}

\maketitle
%Introduction
\section{Introduction}
\label{sec:introduction}
Electrical transport measurements on a \gls{QD} in a magnetic field offer an interesting experimental platform to study the physics in the integer and fractional quantum Hall regime.
Due to the interplay of Coulomb interaction and Landau level quantization, the electron density in the \gls{QD} reconstructs self-consistently into concentric compressible regions separated by incompressible rings \cite{chklovskii_electrostatics_1992, dempsey_electron-electron_1993, evans_coulomb_1993}.
For an internal filling factor $2 > \nu_\mathrm{dot} > 1$, two compressible regions form inside the \gls{QD}, which was studied in seminal experiments measuring the magnetoconductance through the quantum dot \cite{brown_resonant_1989,mceuen_transport_1991,mceuen_self-consistent_1992,mceuen_coulomb_1993}. Each compressible region carries discrete integer charge, and hexagonal phase-diagrams of stable charge as a function of gate voltage and magnetic field were predicted theoretically \cite{evans_coulomb_1993} and measured in different experiments \cite{heinzel_periodic_1994, fuhrer_transport_2001, chen_transport_2009-1, baer_cyclic_2013, sivan_observation_2016-1, liu_electrochemical_2018, roosli_observation_2020}. The width of the incompressible stripe, and thereby the tunneling rate between compressible regions, changes with magnetic field, where experimentally the rate could be lowered sufficiently to measure time-resolved tunneling events inside the \gls{QD} \cite{van_der_vaart_time-resolved_1994,van_der_vaart_time-resolved_1997}.
Electron depopulation in the \gls{QD} alternates between the two compressible regions when the total electron number is reduced one by one \cite{baer_cyclic_2013}.
In the fractional quantum Hall regime, \glspl{QD} exhibit a similar charge reconstruction with incompressible regions assuming a fractional filling factor.
This enables tunneling of fractionally charged quasiparticles between the compressible regions, thereby exhibiting fractional Coulomb blockade \cite{roosli_fractional_2021}, while only electrons can be exchanged with the leads \cite{chang_chiral_2003}.
Opening the tunneling barriers to the regime of weak backscattering, allows for operating these systems as quantum Hall Fabry-Pérot interferometers in the integer \cite{van_wees_observation_1989, taylor_aharonov-bohm_1992, bird_precise_1996, camino_aharonov-bohm_2005-1, godfrey_aharonov-bohm-like_2007, rosenow_influence_2007} and fractional \cite{ofek_role_2010-1, camino_aharonov-bohm_2005, zhou_flux-period_2006, camino_quantum_2007, camino_$e/3$_2007, lin_electron_2009, mcclure_fabry-perot_2012, willett_measurement_2009, willett_alternation_2010, willett_interference_2019, nakamura_aharonovbohm_2019}  quantum Hall regimes, whose behavior is closely related to Coulomb-blockaded \glspl{QD} in magnetic field \cite{zhang_distinct_2009, ofek_role_2010-1, halperin_theory_2011, sivan_observation_2016-1, roosli_observation_2020}. In recent groundbreaking experiments, Fabry-Pérot interferometers in the fractional quantum Hall regime were used to measure anyonic statistics \cite{nakamura_direct_2020-1}, which was also shown in collision experiments using quantum point contacts \cite{bartolomei_fractional_2020}.

% study conductance through QD, explain compressible regions with ref, then go to three main points.
In this paper, we measure magneto-transport through a \SI{1.4}{\micro m}-wide \gls{QD}, and present three different aspects extending and complementing the dot-internal rearrangements of quasiparticles in the integer and fractional quantum Hall regime \cite{van_der_vaart_time-resolved_1994, van_der_vaart_time-resolved_1997, heinzel_periodic_1994, fuhrer_transport_2001, chen_transport_2009-1, baer_cyclic_2013, sivan_observation_2016-1, liu_electrochemical_2018, roosli_observation_2020, roosli_fractional_2021}.
(i) For integer filling factors $6 > \nu_\mathrm{dot} > 2$, the \gls{QD} reconstructs into more than two concentric, compressible regions separated by incompressible stripes \cite{chklovskii_electrostatics_1992,evans_coulomb_1993}. As this allows for various different rearrangements between the multiple compressible regions inside the \gls{QD}, we observe distinct modulations of the conductance resonances, showing a more complex pattern than for two compressible regions at filling factors $2>\nu_\mathrm{dot}>1$ \cite{chen_transport_2009-1, sivan_observation_2016-1, liu_electrochemical_2018, roosli_observation_2020, roosli_fractional_2021}, including changes in  magnetic field period, super-periodicities, and non-periodic features.
(ii) Measuring the conductance through an auxiliary \gls{QD} next to the principal \gls{QD}, we show that internal reconstructions can be detected using an external, capacitively coupled charge detector \cite{field_measurements_1993,buks_dephasing_1998-1,smith_detection_2002,sprinzak_charge_2002,elzerman_few-electron_2003-1,petta_manipulation_2004,ihn_quantum_2009,chen_transport_2009-1,baer_cyclic_2013} in the integer quantum Hall regime.
(iii) We study the temperature dependence of the conductance resonances, which shows a continuous transition between two qualitatively different behaviors when crossing the filling factor range $2 > \nu_\mathrm{dot} > 1$. Additionally, we find differing temperature dependencies of the amplitudes for conductance resonances with different total electron number at $B = \SI{0}{T}$.
Furthermore, we discuss the tuning of the gate voltages, and their influence on the parameters of the \gls{QD}, such as the charging energy, the plunger gate capacitance and the lever arms, as well as their dependence on magnetic field.

\section{Sample and experimental setup}
\label{sec:sample_exp_setup}
% sample
We fabricate the measured sample on a standard modulation doped, single-interface AlGaAs-heterostructure, with a \gls{2DEG} buried \SI{130}{nm} underneath the surface.
A Hall bar is fabricated by wet etching, and we ohmically contact the \gls{2DEG} by annealing AuGeNi contacts. At the temperature $T=\SI{30}{mK}$, we determine the electron mobility $\mu = \SI{5.6e6}{cm^{2}/Vs}$ and the bulk electron density $n_\mathrm{bulk} = \SI{1.44e11}{cm^{-2}}$. A pre-patterned back gate was fabricated approximately \SI{1}{\micro m} below the \gls{2DEG} before overgrowing the heterostructure by molecular beam epitaxy \cite{berl_structured_2016-1}. The bulk electron density and the mobility can be varied by applying a voltage between the \gls{2DEG} and the back gate, which is covering the full area underneath the Hall bar. The back gate was grounded for all measurements presented below, while nevertheless modifying the electron-electron interaction and the electrostatic potential in the sample. The presence of additional gates was used to influence the electrostatic potential in previous experiments \cite{nakamura_aharonovbohm_2019, willett_interference_2019, choi_robust_2015, sivan_interaction-induced_2018}.

\begin{figure*}[tbp]
\includegraphics{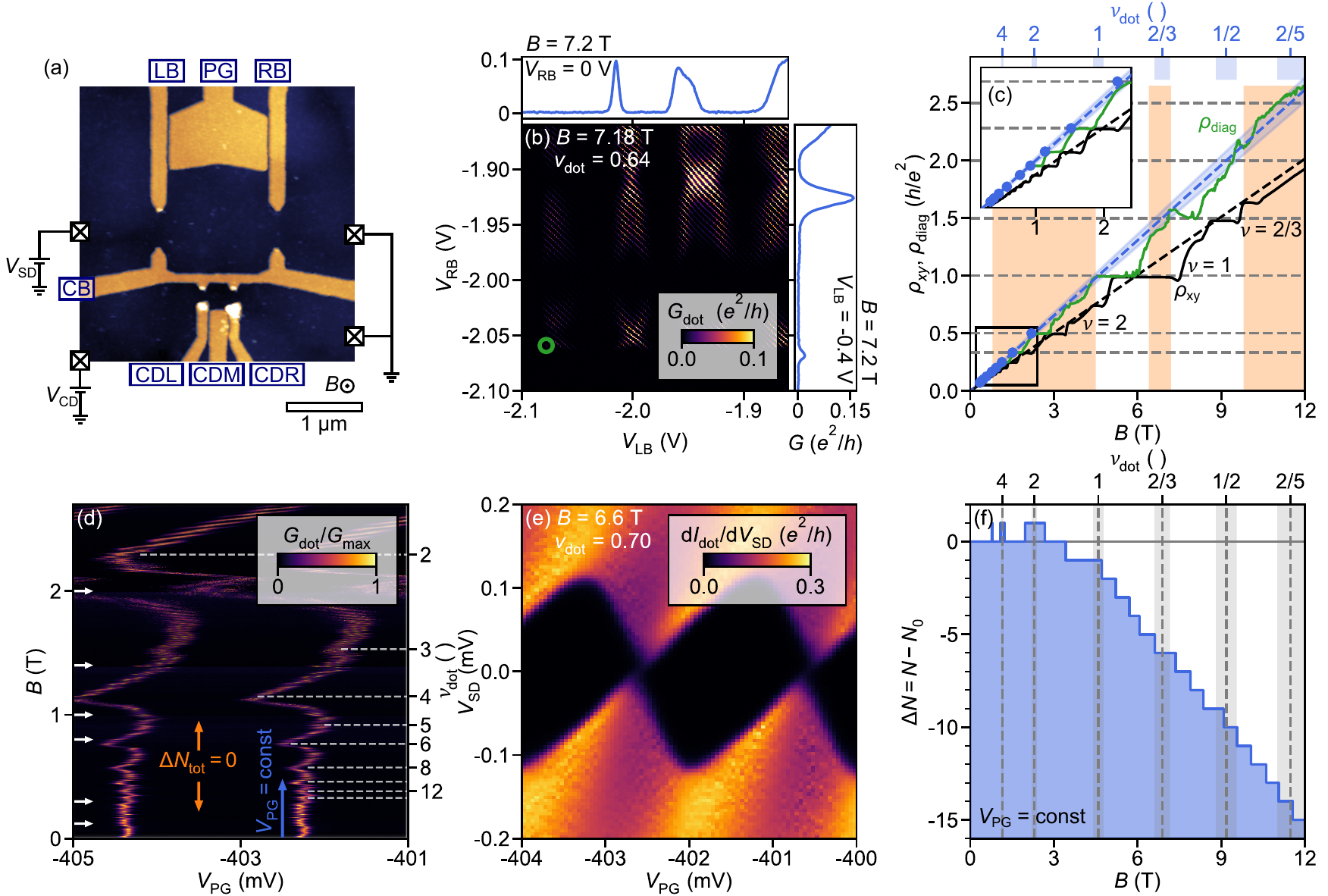}
\caption{(a) Atomic force micrograph of the inner structure of the sample. The metallic gates (yellow) on the sample surface (dark blue) are labeled. Contacts and applied voltages are indicated schematically.
	(b) Conductance $G_\mathrm{dot}$ through the \gls{QD} as a function of the voltages of the left and right barrier gates, $V_\mathrm{LB}$ and $V_\mathrm{RB}$, at $B=\SI{7.18}{T}$. Plots along the axis show the conductance dependence of a single barrier versus the barrier gate voltage $V_\mathrm{LB}$ (or $V_\mathrm{RB}$) while the other barrier gate is grounded (weakly biased), i.e. $V_\mathrm{RB} = \SI{0}{V}$ (or $V_\mathrm{LB}= \SI{-0.4}{V}$) at $B=\SI{7.2}{T}$.
	(c) Transverse resistivities $\rho_\mathrm{xy}$ (black) and $\rho_\mathrm{diag}$ (green, with gates PG and CB defined) as a function of magnetic field. The inset shows a zoom-in on the lower magnetic field region. The blue, dashed line indicates the classical transverse resistivity, expected for the charge density $n_\mathrm{dot}$ as extracted from (d), where blue dots mark 1/B-periodic features. A corresponding filling factor axis $\nu_\mathrm{dot}$ is indicated with the blue shaded regions marking the uncertainty. Periodic modulation of the Coulomb resonances are observed in the orange shaded magnetic field ranges.
	(d) Normalized conductance $G_\mathrm{dot}/G_\mathrm{max}$ as a function of plunger gate voltage and magnetic field (adapted from \cite{roosli_fractional_2021}). Here, $G_\mathrm{max}$ denotes the local maximum over five adjacent traces in plunger gate voltage. We associate the 1/B-periodic modulations in conductance to integer filling factors, and calculate a charge carrier density in the \gls{QD} of  $n_\mathrm{dot} = \SI{1.11 \pm 0.04e11}{cm^{-2}}$. As the tunnel coupling of the barriers changes with magnetic field, we combine seven measurements at different barrier gate voltages, where transitions between individual measurements are indicated by white arrows. They are shifted in plunger gate voltage such that the conductance resonances match at the boundaries. The total electron number $N_\mathrm{tot}$ is constant in-between two resonances (indicated in orange).
	(e) Finite bias measurement of the numerical derivative $dI_\mathrm{dot}/dV_\mathrm{SD}$ as a function of plunger gate voltage and source-drain bias voltage at $B=\SI{6.6}{T}$.
	(f) Change in the total number of electrons  on the \gls{QD}, $\Delta N$, as a function of magnetic field $B$ or filling factor $\nu_\mathrm{dot}$ at constant plunger gate voltage [as indicated by blue arrow in (d)]. } 
\label{fig1}
\end{figure*}

An atomic force microscope (AFM) image in Fig.~\ref{fig1}(a) shows the central gate structure of the quantum dot sample. Lateral metallic top gates (yellow) are deposited directly onto the mesa surface (dark blue) by evaporating gold (Au). Four top gates (LB, PG, RB, and CB) are used to form the \gls{QD} with a lithographic area of $1.4 \times \SI{1.4}{\micro m^2}$, where the lithographic spacing between the barrier gates (between LB and CB, and RB and CB) is \SI{500}{nm}. An additional small quantum dot with a lithographic area of $0.32 \times \SI{0.4}{\micro m^2}$, used for charge detection, can be formed with three additional gates (CDL, CDM, and CDR) next to the primary \gls{QD}, together with the central barrier gate (CB). Both \glspl{QD} are formed by applying a negative voltage between the respective top gates and the \gls{2DEG}, thus depleting the electron gas below the gates at voltages smaller than \SI{-350}{mV}. The parts of the electron gas between gates, that are decoupled from source and drain contacts, are grounded through additional ohmic contacts at the sides of the Hall bar. Thereby, we ensure there are no floating \gls{2DEG} regions which might accumulate charge.
The conductance through the \gls{QD}, $G_\mathrm{dot} = I_\mathrm{dot}/V_\mathrm{SD}$, is calculated by measuring the 2-terminal current $I_\mathrm{dot}$ using an IV-converter, when a voltage $V_\mathrm{SD}$ is applied between the source and drain contacts. Similarly, we determine the 2-terminal conductance through the \gls{CD}, $G_\mathrm{CD} = I_\mathrm{CD}/V_\mathrm{CD}$. While the bias to the primary \gls{QD} and the \gls{CD} is applied through different source contacts, they share the same drain contact. The same sample was used for previous experiments \cite{roosli_fractional_2021}.

% cooldown, temperature, other cooldowns, other samples
The data presented in this paper were taken on one sample during two different cooldowns in a dilution refrigerator. The gate voltages applied for the two different cooldowns are similar, and the tuning of the sample is reproducible. The system reaches the electron temperature $T_\mathrm{e} = \SI{30}{mK}$, and the temperature is varied by applying power to a resistive heater placed next to the mixing chamber. Electron temperatures are determined from the width of the Coulomb resonances of the quantum dot \cite{averin_theory_1991, beenakker_theory_1991}. Changing the density by applying a non-zero voltage to the back gate, we observe similar results (not shown in this paper). Periodic modulations of the conductance resonances were also observed in other samples, and in multiple cooldowns (see \cite{roosli_observation_2020} for a different sample). 

\section{Quantum dot tuning at finite magnetic fields}
\label{sec:QD_tuning}
% form QD
Our experiments are conducted in the Coulomb blockade regime with the conductance of the \gls{QD} being much smaller than $e^2/h$.
The coupling of the \gls{QD} to the leads is controlled by the voltages applied to the center barrier gate (CB), and the left and right barrier gate (LB and RB) forming the left and right tunneling barrier, respectively.
To adjust the coupling to the leads, we change the voltage on the left and right barrier gate (LB and RB), while keeping the gate voltage of the center barrier (CB) constant at around $\SI{-1.2}{V}$.
The plunger gate controls the quantized energy levels in the \gls{QD}, where the plunger gate voltage is usually altered in a small range around \SI{-400}{mV}.
A magnetic field perpendicular to the \gls{2DEG} is applied to the sample.

% localized states in the barriers
In Fig.~\ref{fig1}(b), we measured the \gls{QD} conductance $G_\mathrm{dot}$ as a function of the left and right barrier gate voltages, $V_\mathrm{LB}$ and $V_\mathrm{RB}$, at a magnetic field $B = \SI{7.18}{T}$ ($\nu_\mathrm{dot} = 0.64$). 
Diagonal conductance resonances are observed whenever an energy level in the \gls{QD} is aligned with the Fermi energy of the leads.
The resonances show a superimposed modulation of suppressed conductance along horizontal and vertical stripes.
In the regions where horizontal and vertical conductance stripes cross, the resonances are not suppressed, but broadened and bending instead.
At high magnetic fields, localized states form in the barrier quantum point contacts when the magnetic length is much smaller than the characteristic length scale of the random background potential. Here, we estimate that the length scale of the background potential can be approximated by the spacer thickness \SI{70}{nm}, while the magnetic length at $B=\SI{7.2}{T}$ is $\ell = \sqrt{\hbar /(eB)} \approx \SI{9.6}{nm}$.
We hypothesize that the localized states account for broad resonances in the conductance trough the respective barrier.
The inset above (right of) Fig.~\ref{fig1}(b) shows the conductance through the left (right) barrier as a function of the left (right) barrier gate voltage at $B = \SI{7.2}{T}$, while the other barrier gate is grounded or weakly biased ($V_\mathrm{RB}= \SI{0}{V}$ and $V_\mathrm{LB}= \SI{-0.4}{V}$).
The same voltages are applied to the center barrier (CB) and the plunger gate (PG) as for the measurement in Fig.~\ref{fig1}(b).
We observe broad resonances interrupted by even broader regions of suppressed conductance in-between.
When we measure the conductance through the \gls{QD}, both barriers act in series to the \gls{QD}, therefore modulating the measured conductance in Fig.~\ref{fig1}(b).
The positions of the modulations agree with the individually measured conductance traces of the single barriers in the insets.
The large width of the resonances in the barriers is related to the strong coupling to the leads and a weak lever arm of the barrier gates. The enhanced coupling of the barriers at the crossing of horizontal and vertical stripes leads to a broadening of the Coulomb resonances, while a self-energy shift could account for the bending of the resonances.
We observe this kind of conductance modulations for magnetic fields larger than $\gtrsim \SI{4}{T}$.

\begin{figure}[tbp]
\includegraphics[width=\columnwidth]{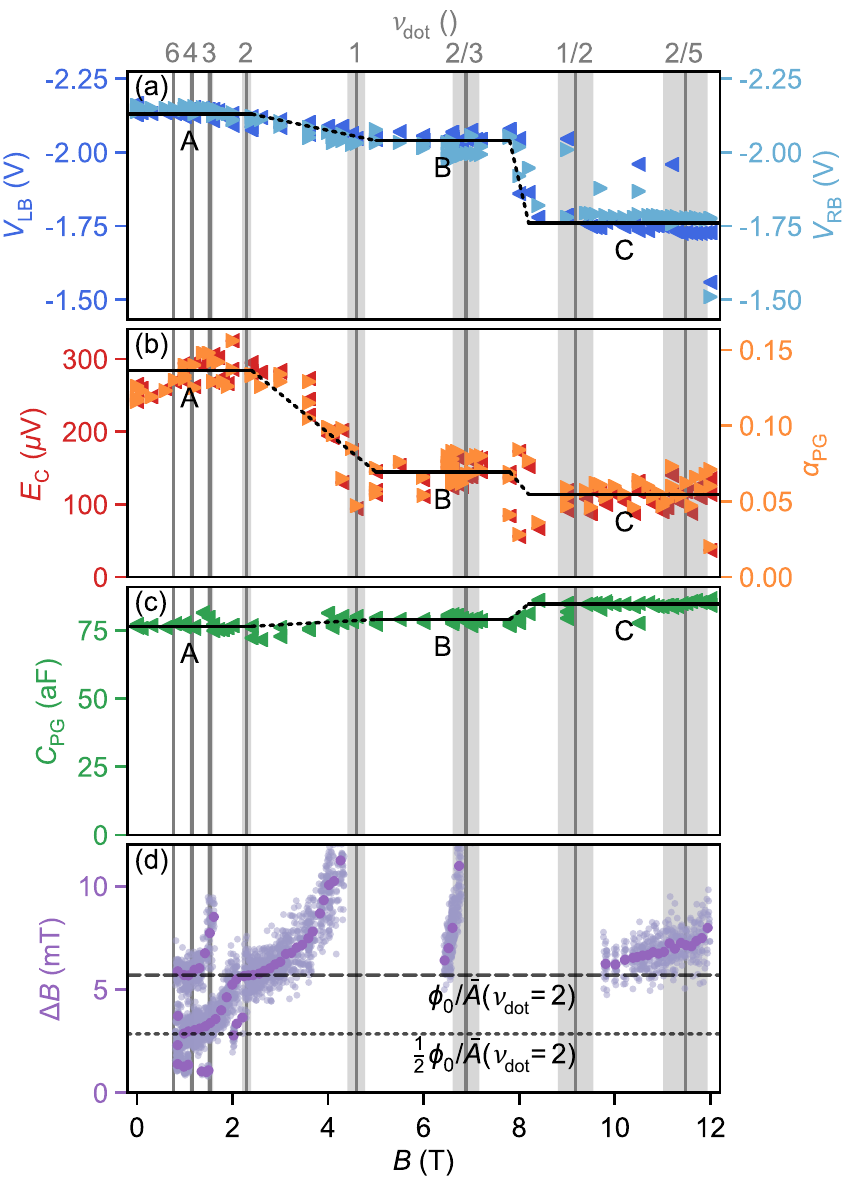}
\caption{Overview of different experimental parameters as a function of the total magnetic field for: (a) right and left barrier gate voltages, $V_\mathrm{RB}$ and $V_\mathrm{LB}$, (b) charging energy $E_\mathrm{C}$ and lever arm $\alpha_\mathrm{PG}$, (c) plunger gate capacitance  $C_\mathrm{PG} = (e^2\alpha_\mathrm{PG})/E_\mathrm{C}$, and (d) the magnetic field period $\Delta B$ (where periodic behavior is observed). The top axis shows the filling factor $\nu_\mathrm{dot}$ in the \gls{QD} with experimental uncertainty (shaded regions). Lines in (a)--(c) act as a guide to the eye, where three different regions are marked. Within one region the experimental parameters are approximately constant with magnetic field. In (d), the light points show the magnetic field spacing $\Delta B$ between two peaks, which are read out manually. The darker points give an average over a range in magnetic field. For integer filling factors, multiple spacings appear as a super-period is visible in the conductance pattern (c.f. Fig.~\ref{fig3}). The dashed (dotted) line indicates a magnetic field spacing corresponding to (half) a flux quantum $\phi_0$ ($\phi_0/2$), assuming the area $\bar{A}$ enclosed by the incompressible ring at internal filling factor $\nu_\mathrm{dot}=2$. }
\label{fig2}
\end{figure}

% tuning barriers
The modulation of the barrier transmission allows for several gate voltage combinations, where the conductance of the barriers is $\ll e^2/h$, i.e., for tuning the \gls{QD} into the weak coupling regime.
For our experiments, we choose the last visible conductance resonance before pinch-off, which corresponds to the most negative barrier gate voltages [marked by a green circle in Fig.~\ref{fig1}(b)].
We aim for Coulomb resonances with a peak conductance of roughly $0.03$ to $0.06~e^2/h$.
The left and right barrier gate voltages used in the measurements are summarized in Fig.~\ref{fig2}(a) as a function of magnetic field (and dot filing factor $\nu_\mathrm{dot}$). Both barrier gate voltages are approximately constant up to \SI{8}{T} (regions A and B) with a slight increase between \SI{2}{T} and \SI{5}{T}, where modulations of the barrier transmission set in.
At \SI{8}{T} the previously used conductance modulation vanishes, such that we adjust the barriers to the next resonance modulation at more positive gate voltage, in order to get Coulomb resonances with the targeted peak conductance for measurements at higher magnetic field (region C).

% bulk density, transverse resitivities rho_xx, rho_diag
In Fig.~\ref{fig1}(c), the transverse bulk resistivity  $\rho_\mathrm{xy}$ is plotted as a function of magnetic field in black. We observe quantized Hall conductance for integer and fractional filling factors. 
The bulk density ($n_\mathrm{bulk} = \SI{1.44e11}{cm^{-2}}$) is determined from fitting the transverse resistivity at low magnetic fields, and the corresponding classical transverse resistivity, $\rho_\mathrm{xy, classical} = B/(e n_\mathrm{bulk})$, is indicated by the black dashed line. 
In the \gls{QD} the density  is reduced compared to the bulk density due to the voltages applied to the top gates. 
The green trace shows the measured transverse resistivity across the \gls{QD}, $\rho_\mathrm{diag}$, where the plunger gate (PG) and the center barrier (CB) are biased with the typical voltages applied in the Coulomb blockade regime, while the barrier gates (LB, RB) are grounded.
The curve is tilted to lower magnetic field for the same resistivity values compared to the bulk measurement, implying lower density. 
While not fully overlapping with the bulk plateaus, quantized conductance is observed for integer filling factors, indicating the presence of quantum Hall states inside the \gls{QD}. 
Signs of fractionally quantized plateaus are present, where the exact quantization is masked by the series resistance of the bulk electron gas. 
A measurement of the transverse resistivity across the fully defined \gls{QD} does not yield information about the density inside the dot, but rather about the density in the barrier region, where it is lowest.

% density inside dot from 1/B-periodic modulations
As described in our previous work \cite{roosli_fractional_2021}, we estimate the density inside the \gls{QD} by measuring the conductance as a function of plunger gate voltage and magnetic field in the integer quantum Hall regime.
In Fig.~\ref{fig1}(d), we show the normalized conductance $G_\mathrm{dot}/G_\mathrm{max}$ measured by sweeping $V_\mathrm{PG}$ and stepping $B$ (adapted from \cite{roosli_fractional_2021}). The normalization $G_\mathrm{max}$ is determined as the maximum conductance observed in five consecutive traces in plunger gate voltage.
As the transmission of the barriers varies with magnetic field, we had to adjust the barrier gate voltages at certain magnetic field values to keep the \gls{QD} weakly coupled to the leads.
Thus, Fig.~\ref{fig1}(d) combines seven blocks of measurements, corresponding to seven magnetic field intervals, with different voltages applied to the barrier gates in each block.
In order to compensate for the resulting shifts of the resonances with changing barrier gate voltages, we shifted the individual blocks such that the conductance resonances match at the transition between blocks (white arrows).

The conductance in Fig.~\ref{fig1}(d) shows Coulomb resonances whenever an energy level of the \gls{QD} is in resonance with the electrochemical potential of the leads, which follow  a $1/B$-periodic modulation in gate voltage (selected extrema indicated by dashed lines).
We associate the individual extremal positions in $V_\mathrm{PG}$ with integer filling factors ($\nu_\mathrm{dot} = 2$, 3, 4, 5, 6, 8, 10, 12 and 14), and estimate the maximum charge carrier density inside the \gls{QD} to $n_\mathrm{dot} = \SI{1.11 \pm 0.04e11}{cm^{-2}}$.
The classical resistivity corresponding to the extracted density $n_\mathrm{dot}$ is plotted in Fig.~\ref{fig1}(c) as a dashed blue line, with the experimental uncertainty indicated by the blue shaded region.
The $1/B$-periodic extrema extracted from the measurement in Fig.~\ref{fig1}(d), are indicated in Fig.~\ref{fig1}(c) as blue points with the theoretical resistivity value $\rho_\mathrm{xy} = (1/\nu) (h/e^2)$ of the associated filling factor $\nu$.
A corresponding filling factor axis with uncertainty (blue shaded) is added at the top of Fig.~\ref{fig1}(c).
We note that the extracted density (blue dashed line, and blue points) corresponds to a lower electron density than the measured transverse resistivity $\rho_\mathrm{diag}$ (green solid line), as the density is further reduced by applying voltages to the barrier gates, in addition to the voltages on the plunger gate (PG) and center barrier (CB).

% finite bias; charging energy and lever arm dependence on magnetic field
Typical Coulomb-blockade diamonds are observed in the finite bias measurement shown in Fig.~\ref{fig1}(e), where we plot the numerically determined differential conductance $dI_\mathrm{dot}/dV_\mathrm{SD}$ as a function of the plunger gate voltage and the source-drain bias at $B = \SI{6.6}{T}$ ($\nu_\mathrm{dot} = 0.70$).
We determine the charging energy $E_\mathrm{C}$ of the \gls{QD} from the extent of the Coulomb-blockade diamonds in $V_\mathrm{SD}$-direction. We further determine the plunger gate lever arm $\alpha_\mathrm{PG}$ connecting a change $\Delta V_\mathrm{PG}$ in plunger gate voltage to the shift $-e\alpha_\mathrm{PG}\Delta V_\mathrm{PG}$ of the electrochemical potential in the \gls{QD}.
In Fig.~\ref{fig2}(b) we plot the charging energy, $E_\mathrm{C}$, and the lever arm of the plunger gate, $\alpha_\mathrm{PG}$, as a function of magnetic field, where both behave similarly.
They are rather constant from $B = \SI{0}{T}$ up to filling factor $\nu_\mathrm{dot} \approx 2$ (with a slight increase from $B = \SI{0}{T}$ to $\nu_\mathrm{dot} \approx 6$; region A).
Between  $\nu_\mathrm{dot} \approx 2$ and $1$, both values continuously decrease and stabilize at roughly half the value attained in the integer quantum Hall regime (region B).
This change is not accompanied by a significant adjustment in barrier gate voltages [see Fig.~\ref{fig2}(b)], nor the other gate voltages.
% Here the reader may ask, why this reduction of the lever arm and the charging energy arises.
We believe that the total dot capacitance $C_\Sigma$ increases as the capacitance of the \gls{QD} to the source and drain regions increases significantly, while the capacitance $C_\mathrm{PG}$ of the plunger gate to the dot remains roughly constant [see below, Fig.~\ref{fig2}(c)]. 
This leads to a reduction of both the charging energy $E_\mathrm{C} = e^2/C_\Sigma$ and the plunger gate lever arm $\alpha_\mathrm{PG} = C_\mathrm{PG}/ C_\Sigma$.
We hypothesize that the increasing capacitance of the dot to the leads is related to the decreasing magnetic length, allowing the states of the dot and the leads to extend further into the barriers, and thereby to couple more strongly to each other.
The temperature dependence of the conductance resonances changes behavior significantly in the same range of magnetic field, as we will discuss later [see Sec.~\ref{sec:temperature_dep}, Fig.~\ref{fig7}(c)--(f)].
At $B \approx \SI{8}{T}$, both values decrease slightly and remain approximately constant for higher magnetic field (region C). This decrease is probably associated with the adjustment of the barrier gate voltages [see Fig.~\ref{fig2}(a)] at the same field, which increases the capacitance between the dot and the leads.

% plunger gate capacitance				
Using the spacing $\Delta V_\mathrm{PG}$ of conductance resonances at vanishing $V_\mathrm{SD}$, we calculate the capacitance $C_\mathrm{PG} = e/\Delta V_\mathrm{PG}$ between \gls{QD} and plunger gate, and plot the resulting values in Fig.~\ref{fig2}(c) as a function of magnetic field. The plunger gate capacitance remains approximately constant for the full range of magnetic field (regions A--C). The slight change at $B \approx \SI{8}{T}$ might be related to the change of the barrier gate voltages [see Fig.~\ref{fig2}(a), region C]. Correspondingly, also the separation of Coulomb resonances $\Delta V_\mathrm{PG}$ is independent of magnetic field.

% change in total electron number in the QD
Following the conductance resonances at constant plunger gate voltage in a measurement like in Fig.~\ref{fig1}(d) over the full magnetic field range [indicated by blue arrow in Fig.~\ref{fig1}(d)], we determine the change of the total electron number in the \gls{QD} as a function of magnetic field.
Whenever a line along the magnetic field axis crosses a charging resonance, the total electron number changes by one. The resulting change of the electron number $\Delta N$ is presented in Fig.~\ref{fig1}(f) in the full magnetic field range from $B = \SI{0}{T}$ to $\SI{12}{T}$.
Here, we assume that the charging energy is roughly independent of the electron number $N$ in the \gls{QD}, and that all the Coulomb resonances are parallel to each other for different occupation numbers, as this analysis combines measurements with different gate voltage settings for different magnetic field ranges, and thus different total occupation numbers $N$.
In the integer quantum Hall regime up to filling factor $\nu_\mathrm{dot} \approx 1$, the total electron number stays constant on average. Beyond filling factor $\nu_\mathrm{dot} \approx 1$, the total electron number continuously decreases with increasing magnetic field, where roughly 10 electrons are expelled from the \gls{QD} within a range of \SI{5}{T}. This is not considered within the simple model presented in previous work \cite{roosli_observation_2020,roosli_fractional_2021}, where no magnetic field dependence of the total electron number is predicted, and we assume this is related to a self-consistent effect.
The total electron number occupying the \gls{QD} cannot be directly extracted from the measurement. From electrostatic simulations, we estimate a total occupation number of $N_0 \approx 800$ to $900$ at $B = \SI{0}{T}$ \cite{roosli_fractional_2021}. As a result, the total number of electrons on the \gls{QD} changes by roughly 2\%  over the range in magnetic field which is studied here.

\section{Dot filling factor $\mathbf{\nu_\mathrm{dot} > 2}$}
\label{sec:higher_integerFF}
\begin{figure*}[tbp]
\includegraphics{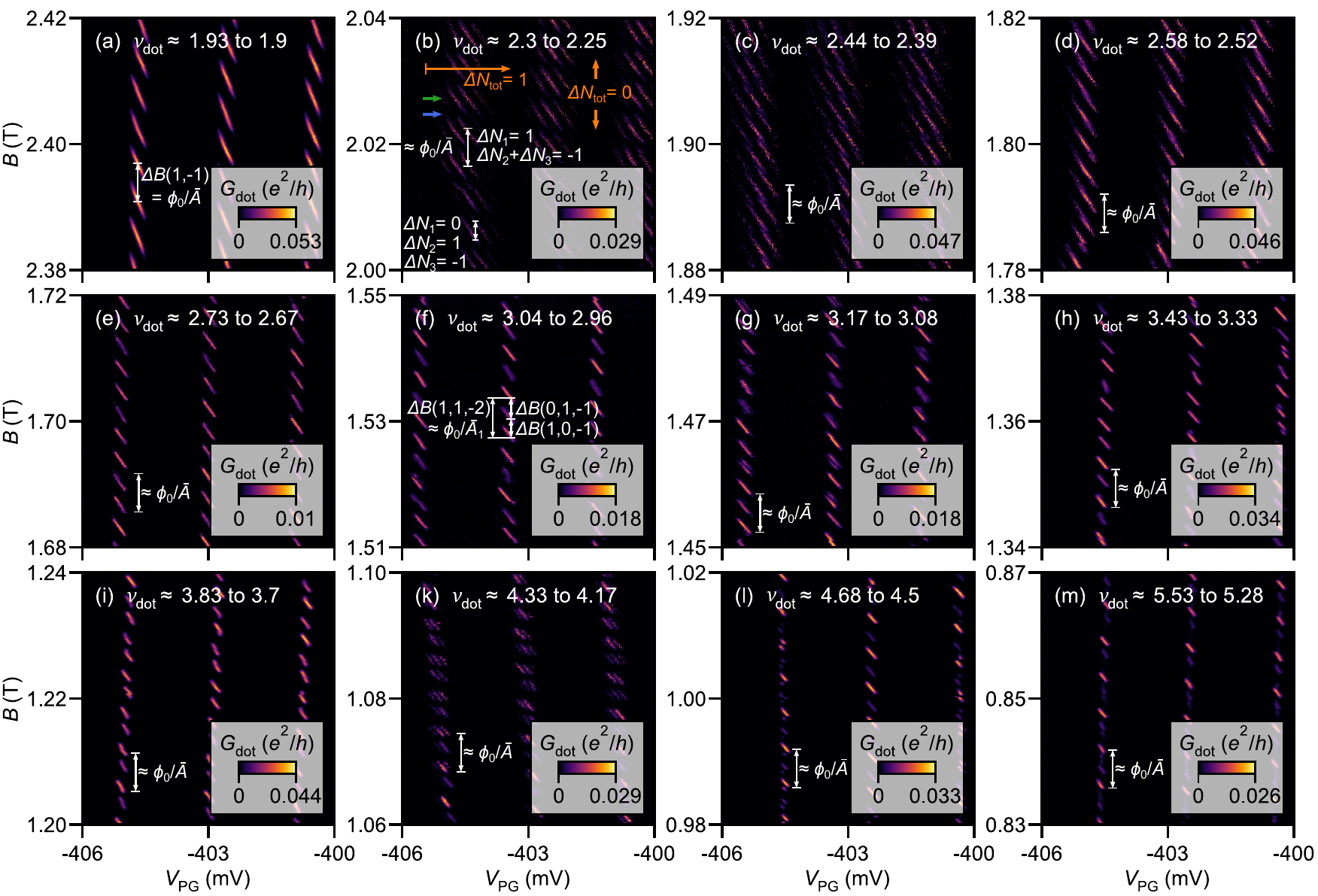}
\caption{Conductance $G_{\mathrm{dot}}$ as function of plunger gate voltage and magnetic field for different magnetic field ranges with dot filling factors $\nu_\mathrm{dot}$ indicated. For (a) $\nu_\mathrm{dot} \lesssim 2$ the conductance shows a periodic behavior, where each shifted resonance corresponds to an internal rearrangement of an electron. For higher filling factors the structure of the conductance resonances gets more intricate, and super-periodic patterns [e.g. in (b), (g), (k), (l) or (m)] and irregularities emerge. The magnetic field period $\Delta B = \phi_0/\bar{A}$ is indicated for $\nu_\mathrm{dot} \lesssim 2$ in (a), and for comparison in (b)--(m), where $\bar{A}$ corresponds to the area enclosed by the incompressible region at $\nu_\mathrm{dot} = 2$. The total electron number changes by one when crossing a Coulomb resonance along plunger gate direction, and remains constant in the region of suppressed conductance in-between resonances [indicated in orange in (b)]. Changes of the electron number on the $i$-th compressible region, $\Delta N_i$, and magnetic field spacings, $\Delta B (\Delta N_i)$, are indicated in (a), (b) and (f).} 
\label{fig3}
\end{figure*}

The rearrangements of quasiparticles inside the \gls{QD} as a function of magnetic field, result in periodic modulations of the conductance resonances [see Fig.~\ref{fig3}(a)], which we observed in wide regions of magnetic field, as indicated by the orange shaded regions in Fig.~\ref{fig1}(c).
The influence of quasiparticle tunneling between two compressible regions on the conductance of a \gls{QD} was analyzed in previous work for filling factors $2 > \nu_\mathrm{dot} > 1$  \cite{van_der_vaart_time-resolved_1994, van_der_vaart_time-resolved_1997, heinzel_periodic_1994, fuhrer_transport_2001, chen_transport_2009-1, baer_cyclic_2013, sivan_observation_2016-1, liu_electrochemical_2018, roosli_observation_2020} and fractional filling factors $\nu_\mathrm{dot} \gtrsim 2/3$, $\nu_\mathrm{dot} \gtrsim 1/3$ \cite{roosli_fractional_2021}. In this section, we discuss the conductance through the quantum dot for integer quantum Hall filling factors greater than two ($\nu_\mathrm{dot} > 2$), where the transport is influenced by the existence of more than two compressible regions inside the \gls{QD}.
Figure~\ref{fig3} shows the \gls{QD} conductance as a function of plunger gate voltage and magnetic field for different filling factors $\nu_\mathrm{dot}$, i.e., centered around different magnetic field values. For all filling factors in Fig.~\ref{fig3}, we observe distinct (pseudo)-periodic patterns resembling the measured conductance at $\nu_\mathrm{dot} \lesssim 2$ shown in Fig.~\ref{fig3}(a). For $\nu_\mathrm{dot} \lesssim 2$ [see Fig.~\ref{fig3}(a)], we observe clear conductance resonances exhibiting a negative slope. The resonances are separated in plunger gate voltage by a region of suppressed conductance, where the \gls{QD} system is in Coulomb blockade and the total charge on the \gls{QD} remains constant. Whenever the change in magnetic field causes a rearrangement of an electron between the two compressible regions in the \gls{QD}, this results in a shift of the resonances in plunger gate voltage  \cite{roosli_observation_2020}.

\begin{figure}[tbp]
\includegraphics[width=\columnwidth]{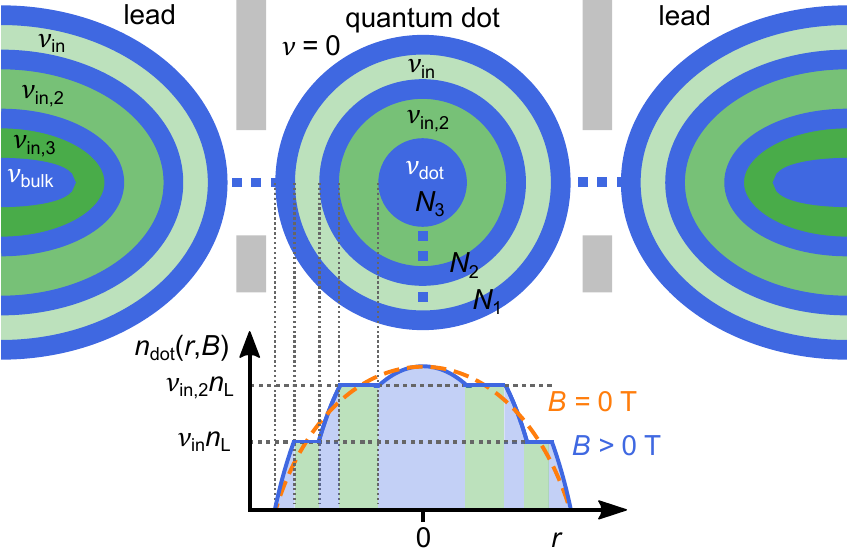}
\caption{Schematic of the \gls{QD} system for the integer quantum Hall regime at filling factor  $3 > \nu_\mathrm{dot} > 2$, where three concentric compressible regions (blue) are separated by incompressible rings at filling factors $\nu_{\mathrm{in,}i}$ (green). Each compressible region hosts an integer number of electrons $N_i$. Compressible regions are tunnel coupled to their next-nearest neighbours, while only the outermost compressible region is weakly coupled to the source and drain leads over controllable tunnel barriers. In general, the filling factor in the leads is larger than inside the \gls{QD}, as the density is reduced by the gate  voltages applied to define the dot. This may result in additional compressible regions, which does not influence the experiment as the conductance is mostly influenced by the dot internal charge rearrangements. We schematically show the radial density distribution in the \gls{QD} for $B = \SI{0}{T}$ (orange, dashed) and $B > \SI{0}{T}$ (blue). } 
\label{fig3b}
\end{figure}

% transition from filling factor 2 to 3
\textbf{Conductance evolution for filling factor \boldmath$\nu_\mathrm{dot}$ from 2 to 3.}
When increasing the filling factor inside the \gls{QD} above two, an additional third compressible region forms in the center of the \gls{QD}, as the next Landau level is pushed below the Fermi energy, thereby partially filling its states \cite{chklovskii_electrostatics_1992, dempsey_electron-electron_1993, evans_coulomb_1993}. This compressible region contains a discrete integer number of charge carriers $N_i$, and is separated from the next compressible region by an incompressible ring.
The schematic in Fig.~\ref{fig3b} shows the reconstruction of the charge density into three concentric compressible regions (blue), which are insulated by incompressible rings (light green).
Only the outermost compressible ring is weakly coupled to the leads over tunnel barriers, which are tuned by the barrier gate voltages, while we assume that direct conductance through all inner regions is suppressed.
The compressible regions are tunnel coupled to their neighboring compressible regions, respectively. The density in the \gls{QD} is reduced compared to the bulk as a result of the applied gate voltages, which can lead to a reconstruction of the bulk into additional compressible and incompressible regions.
However, in our experiments investigating the dot internal rearrangements, only the internal reconstruction and the filling factor $\nu_\mathrm{dot}$ inside the \gls{QD} are important, while the bulk reconstruction might differ.

For $\nu_\mathrm{dot} \gtrsim 2$ in Fig.~\ref{fig3}(b) and (c), the conductance shows bunches of multiple peaks along the magnetic field direction [see Fig.~\ref{fig3}(b), green arrow], which are interrupted by regions of weak conductance [see Fig.~\ref{fig3}(b), blue arrow]. In plunger gate voltage direction, the resonances are separated by regions of suppressed conductance, where the total charge of the \gls{QD} is constant, i.e., $\Delta N_\mathrm{tot} = \mathrm{const}$ [indicated by orange arrow in Fig.~\ref{fig3}(b)]. 
Changing the magnetic field at constant plunger gate voltage causes redistributions of electrons between the three compressible regions inside the \gls{QD}. Due to Coulomb interaction between compressible regions, the corresponding conductance resonances are shifted in plunger gate voltage, and form bunches of conductance peaks [indicated by green arrow in Fig.~\ref{fig3}(b)].
Thereby, we associate the different peaks within one bunch to rearrangements between the inner two compressible regions, while the charge on the outermost region remains unchanged, i.e., charging processes with $\Delta N_1 = 0$, $\Delta N_2 = 1$, and $\Delta N_3 = -1$ [as indicated in Fig.~\ref{fig3}(b)].
When switching to a different bunch of peaks in magnetic field direction, an electron charge is rearranged between the outermost compressible region and the inner two regions, i.e., $\Delta N_1 = 1$, $\Delta N_2 +\Delta N_3 = -1$. This is supported by the observation that the resonance bunches exhibit the same magnetic field periodicity as the rearrangements of the resonances at  $\nu_\mathrm{dot} \lesssim 2 $ in Fig.~\ref{fig3}(a), indicated as $\phi_0/\bar{A}$. This period corresponds to a change of the flux through the \gls{QD} by one flux quantum $\phi_0$, which results in a change of charge on the outermost compressible region by the elementary charge $e$.

The tunneling rate between compressible regions is determined by the widths of the incompressible regions, where time-resolved tunneling could be observed previously \cite{van_der_vaart_time-resolved_1994, van_der_vaart_time-resolved_1997}.
We believe that for Fig.~\ref{fig3}(b)--(d) the tunneling rate of electrons between compressible regions is comparable to the measurement time scale. 
As a result, we observe discrete jumps in the conductance which are perceived as `noisiness' in the measurement.
The width of the incompressible regions shrinks with decreasing magnetic field \cite{van_der_vaart_time-resolved_1994,van_der_vaart_time-resolved_1997}, which leads to faster tunneling rates between the compressible regions.
Consequently, fewer switching events are observed in the conductance for lower magnetic fields when comparing Fig.~\ref{fig3}(b)--(f).
For $\nu_{\mathrm{dot}} \lesssim 2$ (but close to filling factor 2), no  discrete jumps are observed in the conductance in Fig.~\ref{fig3}(a). For $\nu_{\mathrm{dot}} \gtrsim 2$, the width of the incompressible ring between the two outermost compressible regions further decreases with increasing filling factor, resulting in an increased tunnel coupling between those two regions in comparison to filling factors $\nu_{\mathrm{dot}} \lesssim 2$.
Conversely, for filling factors $\nu_{\mathrm{dot}} \gtrsim 2$, the innermost compressible region is relatively small, and separated from the middle compressible region by a comparably wide incompressible stripe, which results in a slow tunneling rate compared to the measurement time scale.
This provides evidence that the discrete conductance jumps between peaks within one bunch in Fig.~\ref{fig3}(b) and (c), result from tunneling events between the inner and the middle compressible regions, not involving tunneling processes to the outermost compressible region.
Similar observations were previously reported for the regime at filling factor $2 \gtrsim \nu_{\mathrm{dot}} \gtrsim 1$  \cite{van_der_vaart_time-resolved_1994,van_der_vaart_time-resolved_1997,roosli_observation_2020}.

When increasing the filling factor further towards $\nu_\mathrm{dot} = 3$ for Fig.~\ref{fig3}(d)--(f), the bunching of resonances vanishes, while the shifted conductance resonances due to internal charge rearrangements persist and get more regular. The magnetic field period in Fig.~\ref{fig3}(e) equals roughly half the period for $\nu_\mathrm{dot} \lesssim 2$ in Fig.~\ref{fig3}(a), which corresponds to half a flux quantum $\phi_0/2$, when we assume a similar area enclosed by the outermost incompressible region for both filling factor regimes.
Upon addition of a flux quantum to the flux threading the area of the \gls{QD} by increasing the magnetic field, an electron is rearranged from the innermost compressible region to each outer compressible region.
Therefore, two electrons from the innermost region are rearranged per flux quantum $\phi_0$ for filling factors $\nu_\mathrm{dot} \lesssim 3$. 
Consequently, we conjecture that the rearrangements of the resonances observed in Fig.~\ref{fig3}(d)--(f), correspond to the alternating cyclic population of the outermost and middle compressible region by electrons tunneling from the innermost compressible region.
In previous experiments, cyclic depopulation of two Landau levels in a quantum dot close to filling factor 2 was associated with alternating height of the resonance amplitudes \cite{baer_cyclic_2013}.
Other works report magnetic field periods corresponding to half a flux quantum $\phi_0/2$ around filling factor $5/2$ for Fabry-Pérot interferometers \cite{choi_robust_2015, sivan_interaction-induced_2018} or capacitance measurements of a \gls{QD} \cite{demir_correlated_2020}, where this is accompanied by a tunneling quasiparticle charge of $2e$, indicating electron pairing \cite{frigeri_electron_2020}.

% evolution for 3 to 4
\textbf{Conductance evolution for filling factor \boldmath$\nu_\mathrm{dot}$ from 3 to 4.}
For dot internal filling factors $4 > \nu_\mathrm{dot} > 3$, the \gls{QD} reconstructs into four concentric compressible regions, which host an integer number of charge carriers, and are separated by incompressible rings (similar to schematic in Fig.~\ref{fig3b} with an additional compressible ring).
The conductance resonances as function of the plunger gate and magnetic field shown in Fig.~\ref{fig3}(f)--(i) exhibit a mostly periodic pattern with some irregularities. We associate the different resonances with discrete charge states of the four compressible regions. 
The resonance amplitudes show variations for different charge reconstructions, and we observe closely spaced double peak features for some resonances [see Fig.~\ref{fig3}(g) and (h)].
In addition, some resonances follow different slopes, while others show a bending, deviating from a straight line [see Fig.~\ref{fig3}(h) and (i)]. 
In Fig.~\ref{fig3}(i), we observe a slight grouping of peaks into pairs. 
The observed variations on a resonance line of the same total charge inside the \gls{QD} closely resemble the variations on next-neighbouring resonances with a total charge different by $\pm e$, but are shifted in magnetic field direction. 

% evolution from 4 to 6
\textbf{Conductance evolution for filling factor \boldmath$\nu_\mathrm{dot}$ from 4 to 6.}
Increasing the filling factor beyond $\nu_\mathrm{dot} > 4$, as shown in Fig.~\ref{fig3}(k), we observe again a bunching of resonances in magnetic field direction, similar to $\nu_\mathrm{dot} \gtrsim 2$ [see Fig.~\ref{fig3}(b) and (c)]. 
Contrastingly, here the resonances within one bunch are spaced more closely in plunger gate voltage compared to the previous case.
In magnetic field, the bunches are periodically spaced by roughly half the period at $\nu_\mathrm{dot} \lesssim 2$ [see Fig.~\ref{fig3}(a)], which corresponds to half a flux quantum $\phi_0/2$, assuming a similar area enclosed by the outermost incompressible ring.
Additionally, a slight pairing of bunches is observed, where the periodicity of the pairs roughly corresponds to the magnetic field period for $\nu_\mathrm{dot} \lesssim 2$ in Fig.~\ref{fig3}(a), i.e. a flux quantum $\phi_0$.
The `noisiness' of the conductance suggests slow intra-dot tunneling rates compared to the measurement times, at least for some of the intra-dot processes, as discussed above for filling factors $3 > \nu_\mathrm{dot} > 2$ [c.f. Fig.~\ref{fig3}(b)--(f)].

For higher filling factors in Fig.~\ref{fig3}(l) and (m) no clear bunching of resonances is observed, while the amplitude of different resonances exhibits variations.
We interpret this as an indication that certain inner compressible regions, different from the outermost region, are directly tunnel-coupled to the leads.
In addition, closely spaced double peaks and other irregularities are visible. The observed pattern is not repeated on the next resonance with the total electron number changed by one (as previously observed for $4 > \nu_\mathrm{dot} > 3$ with a shift in magnetic field).
In particular for Fig.~\ref{fig3}(l), the left resonance shows closely spaced double peak features, which are not observed on the middle resonance, whereas they are observed for the right resonance.

The increasing number of compressible regions allows for various internal charge rearrangements between the different regions, resulting in a conductance resonance pattern that is more intricate compared to the observed pattern at $\nu_\mathrm{dot} \lesssim 2$ for two compressible regions [see Fig.~\ref{fig3}(a)].
The periodic cyclic population, which is observed for lower filling factors, might be broken by the differences in the areas enclosed by the incompressible rings at different filling factors $\nu_{\mathrm{in}, i}$. 
Thereby, the effective magnetic field change required to alter the flux through an incompressible region by one flux quantum $\phi_0$,  differs for the individual incompressible regions.
Such a change of flux results in a rearrangement of $\nu_{\mathrm{in}, i}$ electrons between the neighbouring outer and inner compressible region of the corresponding incompressible ring at filling factor $\nu_{\mathrm{in}, i}$.
As a consequence of the different areas, the periodicities for the charge rearrangements vary for the individual compressible regions.
The differences in area are expected to be more pronounced for higher filling factors, and close to the depopulation of the respective innermost compressible region.
Furthermore, we expect the time-scales of the internal rearrangements to be dominated by the Landau level gaps, which are comparably larger than the Zeeman gap for the studied range of magnetic field. We hypothesize that this could be related to the bunching of resonances observed at larger dot internal filling factors.

\textbf{Full magnetic field spacing evolution with filling factor \boldmath$\nu_\mathrm{dot}$.}
In Fig.~\ref{fig2}(d) we plot the magnetic field spacing $\Delta B$ of resonances at constant plunger gate voltage, as a function of the total magnetic field $B$ and dot internal filling factor $\nu_{\mathrm{dot}}$.
Each light purple point corresponds to the spacing in magnetic field between two (neighboring) conductance resonances, which is extracted by manually fixing the peak positions in conductance maps as a function of plunger gate voltage and magnetic field [as shown in Fig.~\ref{fig3}(a)--(m)].
Dark points represent the average magnetic field spacing over a certain range in magnetic field.
The dashed (dotted) line denotes the magnetic field period associated with (half) a flux quantum $\phi_0$ ($\phi_0/2$) at filling factor $\nu_\mathrm{dot} = 2$ assuming an incompressible area $\bar{A}(\nu_\mathrm{dot} = 2)$.
For the higher filling factors in the integer quantum Hall regime, we observe super-periodic and irregular features in the position and amplitude of the conductance resonances [see Fig.~\ref{fig3}(a)--(m) and discussion above]. 
The magnetic field spacings between those features are also determined and analysed by the manual extraction of the magnetic field spacing in Fig.~\ref{fig2}(d), which can result in multiple different values of the spacing around the same total magnetic field.
In this regime, multiple average values are formed by including only similar spacings. 

We  have described the evolution of the magnetic field period for filling factors  $2>\nu_\mathrm{dot}$ in detail in our previous work \cite{roosli_fractional_2021}.
For filling factors $2 > \nu_\mathrm{dot} > 1$, we observe a magnetic field period that corresponds to a flux quantum $\phi_0$. With decreasing filling factor and increasing field, the magnetic field period, $\Delta B =\phi_0/\bar{A}$, increases as the area $\bar{A}$, enclosed by the incompressible region at filling factor $\nu_\mathrm{in} = 1$, decreases. It eventually diverges at $\nu_\mathrm{dot} = 1$ as the inner compressible region in the \gls{QD} disappears. A similar behaviour is observed in the fractional quantum Hall regime at $\nu_\mathrm{dot} \gtrsim 2/3$ and $\nu_\mathrm{dot} \gtrsim 1/3$, where the magnetic field period corresponds to $\phi_0/2$ and $\phi_0$, respectively \cite{roosli_fractional_2021}.

Now, we consider the magnetic field spacing at higher integer filling factors $6 > \nu_\mathrm{dot} > 2$.
For filling factors $4 > \nu_\mathrm{dot} > 3$, we observe a clear period that can be associated to half a flux quantum, when assuming a similar area enclosed by the incompressible regions as for $\nu_\mathrm{dot} = 2$. Reminiscent of the behaviour at $2 > \nu_\mathrm{dot} > 1$, the period increases with decreasing filling factor, and converges towards one flux quantum when approaching $\nu_\mathrm{dot} = 2$.
An additional smaller period is visible close to $\nu_\mathrm{dot} = 2$, which corresponds to the bunching of conductance resonances observed in this regime.
Further additional periods are observed for $6 > \nu_\mathrm{dot} > 4$, where one takes a value close to the period at $\nu_\mathrm{dot} = 2$ and seems to diverge with decreasing filling factor. On the other hand, we oberve periods that are close to a quarter of the period at $\nu_\mathrm{dot} = 2$.
Assuming cyclic depopulation, where the innermost region loses an electron to each remaining compressible region upon addition of a flux quantum, we expect to observe a magnetic field period corresponding to $1/(M-1)$ of a flux quantum close to the integer filling factors, with $M = \left \lceil{\nu_\mathrm{dot}}\right \rceil$ denoting the number of involved Landau levels.
While we observe indications of higher harmonic magnetic field periods, this argument is too simplified for our \gls{QD}-system, and a detailed model considering the differences in the areas enclosed by the incompressible regions and the different couplings between the compressible regions is needed.

From the evolution of the magnetic field periodicity [see Fig.~\ref{fig2}(d)], we infer that the area(s) enclosed by the incompressible region(s) and their evolution, do not directly influence the charging energy $E_\mathrm{C}$, the lever arm $\alpha_\mathrm{PG}$ or the plunger gate capacitance $C_\mathrm{PG}$ of the \gls{QD} plotted in Fig.~\ref{fig2}(a)--(c). This is consistent with the reconstruction of the zero-field density distribution at high magnetic fields \cite{roosli_fractional_2021}.

In summary, we observe a modulation of the Coulomb resonances of the \gls{QD} due to the internal reconstruction of the Landau levels for filling factors $6 > \nu_\mathrm{dot} > 2$.
Compared to $2 \gtrsim \nu_{\mathrm{dot}}$, where two compressible regions exist inside the \gls{QD}, we observe rich patterns of the conductance resonances with different periodicities, including super-periodic and pseudo-periodic features.
We relate this to the more intricate internal reconstruction of the \gls{QD} and the interplay of the various different compressible regions.
We find reduced magnetic field periods for higher integer quantum Hall states.
Cyclic depopulation of the inner compressible region, where an electron is rearranged from the innermost to each outer compressible region upon a change of one flux quantum, results in an reduced periodicity, which however cannot completely explain the observed behavior. 
Understanding internal reconstruction and quasiparticle tunneling at higher integer filling factors might provide valuable insight for studying the reconstruction of more complex fractional quantum Hall states in \glspl{QD} in the future.

\section{Charge detection}
\label{sec:charge_detection}
In this section, we employ electrostatic charge detection techniques to examine the charge state of the \gls{QD}, thereby complementing the direct measurement of the conductance through the \gls{QD}. For charge sensing we use the conductance resonance of an additional smaller \gls{QD} [in the following referred to as \glsfirst{CD}] next to the primary \gls{QD}  \cite{field_measurements_1993, buks_dephasing_1998-1, smith_detection_2002, sprinzak_charge_2002, elzerman_few-electron_2003-1, petta_manipulation_2004, ihn_quantum_2009, chen_transport_2009-1, baer_cyclic_2013}. When the charge state of the primary dot changes, the resonance of the \gls{CD} is shifted due to the capacitive coupling between the two \glspl{QD}, and as consequence we observe a step in the conductance of the \gls{CD}. The \gls{CD} is formed by applying negative voltages to the gates labelled charge detector left (CDL), charge detector middle (CDM), and charge detector right (CDR), as well as the center barrier (CB) [see Fig.~\ref{fig1}(a)], and tuned to the steep edge of a Coulomb resonance in the few electron regime. A bias voltage $V_\mathrm{CD}$ is applied to the source contact over an IV-converter, while the drain contact is grounded, and the resulting current $I_\mathrm{CD}$ is measured. The same drain contact is used for the \gls{QD} and the \gls{CD}, where we carefully calibrate the offset voltages.

\begin{figure}[tbp]
\includegraphics[width=\columnwidth]{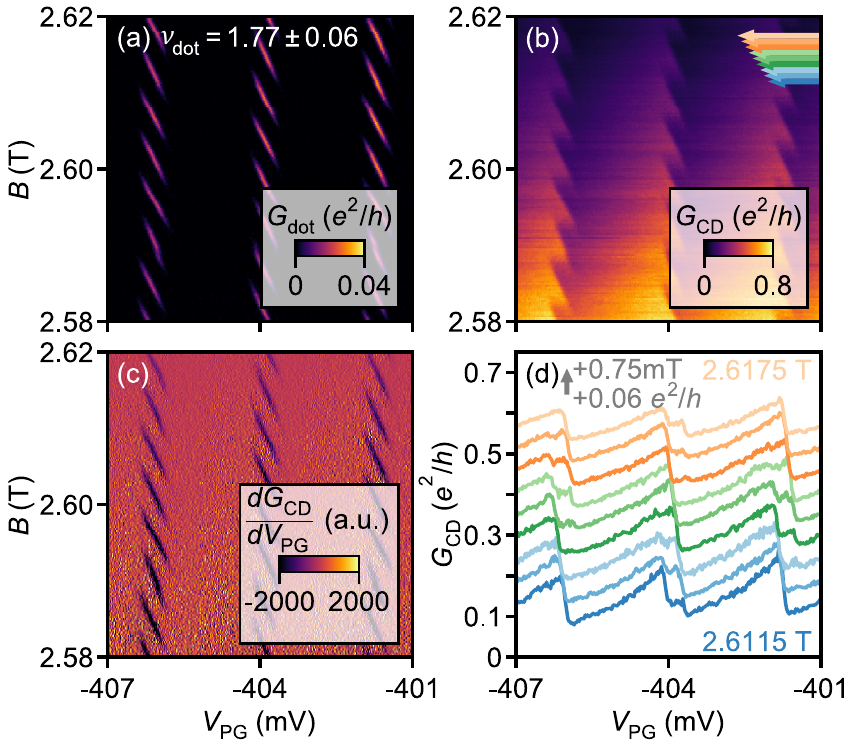}
\caption{Comparison of conductance through the \gls{QD}, $G_\mathrm{dot}$, and \glsfirst{CD} conductance, $G_\mathrm{CD}$, around filling factor $\nu_\mathrm{dot} \approx 1.77$: (a) \gls{QD} conductance $G_\mathrm{dot}$ as a function of plunger gate voltage and magnetic field; (b) conductance $G_\mathrm{CD}$ of the \gls{CD} for the same measurement as in (a). (c) Numerical derivative of the \gls{CD} conductance $dG_\mathrm{CD}/dV_\mathrm{PG}$ shown in (b) in direction of the plunger gate voltage. (d) Conductance $G_\mathrm{CD}$ as a function of the plunger gate voltage, representing cuts along fixed magnetic field values indicated in (b). Different traces are separated in magnetic field by $\SI{+0.75}{mT}$, and an offset of $+0.06 e^2/h$ is added in conductance for clarity. The \gls{CD} shows clear conductance steps where the the charge state of the outer region changes, while charge rearrangements within the dot are not observed. The same charging lines are observed for the \gls{CD} data as for the direct conductance through the \gls{QD}.} 
\label{fig4}
\end{figure}

Figure~\ref{fig4} shows the direct comparison of the conductance through the primary \gls{QD} in Fig.~\ref{fig4}(a) and the charge detector conductance in Fig.~\ref{fig4}(b), as a function of plunger gate voltage and magnetic field around $B = \SI{2.6}{T}$, corresponding to $\nu_\mathrm{dot} = 1.77$.
We find clear steps in the \gls{CD} conductance $G_\mathrm{CD}$ whenever a resonance is observed in the direct dot conductance $G_\mathrm{dot}$.
In Fig.~\ref{fig4}(d), line cuts of the \gls{CD} conductance are plotted at different fixed magnetic field values indicated in (b) as a function of the plunger gate voltage.
The charging of one electron corresponds to a change of the \gls{CD} signal by roughly 50\% of the maximal conductance, which indicates a good \gls{CD} sensitivity.
Charging events detected with the \gls{CD} are more clearly visible in the numerical derivative $dG_\mathrm{CD}/dV_\mathrm{PG}$, which is shown in Fig.~\ref{fig4}(c) for the measurement in Fig.~\ref{fig4}(b).
Employing the \gls{CD}, we only observe the charging lines of the outer compressible region.
For charging the inner region, the switching does not occur immediately at the corresponding boundary of the charge stability diagram and is spread out in Fig.~\ref{fig4}(b)~and~(c) in between the visible resonances or steps, respectively.
The inter-dot charge rearrangement lines are not visible in the \gls{CD} conductance.
The charge distributions, where an additional electron is either on the inner or the outer compressible region, are both centered on the \gls{QD}, and the difference in the electric potential at the \gls{CD} is negligibly small.
Detecting this transition might be possible with a more sensitive \gls{CD}, i.e. either having a higher capacitive coupling or a steeper conductance resonance.
In general, the charge sensing signal agrees with the direct conductance measured through the dot, while not rendering any additional charging lines visible. 

For certain magnetic fields, the trace of the conductance in Fig.~\ref{fig4}(a) shows double peaks as a function of plunger gate voltage, where each individual peak corresponds to a different resonance line with the same total charge $N = N_1 + N_2$ on the \gls{QD}, while the number of electrons $N_i$ on the inner ($i=2$) and outer ($i = 1$) compressible regions differs [i.e., the two charge states $(N_1, N_2)$ and $(N_1+1, N_2-1)$] \cite{van_der_vaart_time-resolved_1994, van_der_vaart_time-resolved_1997, heinzel_periodic_1994, fuhrer_transport_2001, chen_transport_2009-1, baer_cyclic_2013, sivan_observation_2016-1, liu_electrochemical_2018, roosli_observation_2020}.
The measured conductance is an average over the two charge states in the \gls{QD} that follows a thermally activated Boltzmann distribution \cite{van_der_vaart_time-resolved_1997}.
This can also be observed for the \gls{CD} conductance in Fig.~\ref{fig4}(d), where at a fixed magnetic field a double step is found.
The two steps correspond to the two different internal charge states, and are of different height due to the averaging over the probability distribution.
The sum of the two jumps adds up to charging the \gls{QD} with one full electron, as the total electron number on the dot changes by one when crossing a band of resonances.
This provides a direct indication that a single charging line is split up into two lines with effective charges $q_1,~q_2<e$ and $q_1+q_2 = e$,  where $q_i = ep_i$ with the probabilities $p_i$ ($i=1,2$) given by the thermally activated Boltzmann distribution \cite{van_der_vaart_time-resolved_1997}.

% fractional regime CD
In the fractional quantum Hall regime, it was not possible to use the charge detector as a probe for the charge state of the \gls{QD}. 
We assume that the localized states, which form close to the \gls{CD} gates (CDL, CDM, CDR, and CB) at high magnetic field (c.f. Sec.~\ref{sec:QD_tuning} about \gls{QD} tuning), interfere with the \gls{CD}.
The lithographic spacing of the \gls{CD} gates (lithographic area of $0.32 \times \SI{0.4}{\micro m^2}$ and barrier spacing of \SI{200}{nm}) is too small to allow for controlled tuning of the \gls{QD} in the few electron regime at high magnetic fields.
We tried using the resonances of random localized states, but could not reach sufficient detection sensitivities, mainly due to the missing steepness of the conductance resonance as a function of the gate voltage.
We see no intrinsic problem with the charge detection for fractional filling factors, and think that an improved gate design customized for high magnetic fields would offer new possibilities.
However, the  tendency of the density profile to stay close to the distributions at $B = \SI{0}{T}$, may make detecting reconstructions extremely hard.
A solution is to use the outermost compressible region of the primary \gls{QD} as a charge detector for the dot internal rearrangements, as was presented in the direct transport experiments above and shown in previous works \cite{van_der_vaart_time-resolved_1994, van_der_vaart_time-resolved_1997, heinzel_periodic_1994, fuhrer_transport_2001, chen_transport_2009-1, baer_cyclic_2013, sivan_observation_2016-1, liu_electrochemical_2018, roosli_observation_2020, roosli_fractional_2021}.

\section{Temperature dependence}
\label{sec:temperature_dep}
\begin{figure}[tbp]
\includegraphics[width=\columnwidth]{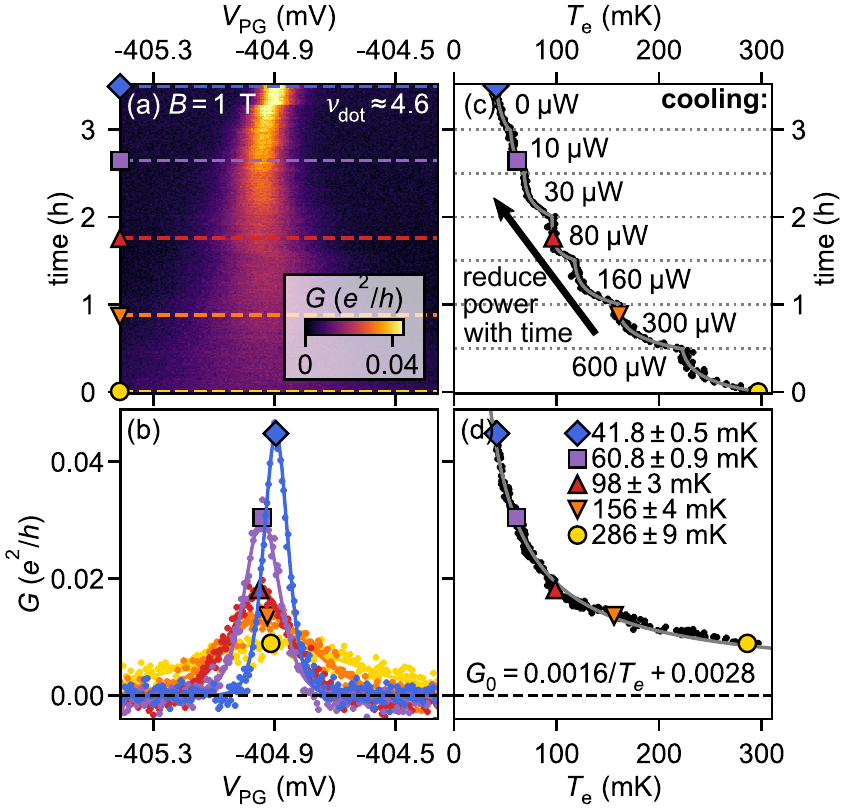}
\caption{Temperature dependence at a fixed magnetic field $B = \SI{1}{T}$ ($\nu_\mathrm{dot} = 4.6 \pm 0.2$):
	(a) The conductance $G_\mathrm{dot}$ is measured as a function of plunger gate voltage and time, combining seven measurements. At the beginning of each measurement, the applied heating power at the mixing chamber, $P_\mathrm{MC}$, is set to a constant value indicated in (c). Before the first measurement the system is completely equilibrated at $P_\mathrm{MC} = \SI{1000}{\micro W}$. During each measurement the \gls{QD} thermalizes at constant applied heating power, and the electron temperature $T_\mathrm{e}$ decreases continuously.
	(b) Conductance shown as a function of plunger gate for different times indicated in (a) [dashed lines]. The electron temperature is determined by a least square fit of the peak shape for single level transport in Eq.~(\ref{eq:single_level_temp}).
	In (c) the resulting electron temperature $T_\mathrm{e}$ is plotted as a function of time. The seven consecutive measurements are separated by dotted lines and the respective power $P_\mathrm{MC}$ is indicated. For each measurement the temperatures are fitted with an exponential curve (gray solid line). The fits are used as a calibration of electron temperature $T_\mathrm{e}$ as a function of time in later measurements (employing the same heating protocol).
	(d) shows the peak amplitude $G_0$ as a function of the electron temperature, where the solid gray line is a fit with a $1/T_\mathrm{e}$-dependence characteristic for single level transport. } 
\label{fig5}
\end{figure}

Now, we study the temperature dependence of the conductance resonances and how it evolves for different filling factors inside the \gls{QD}.
First, we analyze the temperature dependence at a fixed magnetic field and explain the measurement procedure.
Fig.~\ref{fig5}(a) shows the conductance as a function of plunger gate voltage and time for a magnetic field $B = \SI{1}{T}$ corresponding to the filling factor $\nu_\mathrm{dot} = 4.6 \pm 0.2$.
We combine seven consecutive measurements of \SI{30}{min} duration each, where a constant heating power is applied to the mixing chamber during each measurement.
Before the first measurement, a heating power of $P_\mathrm{MC} = \SI{1000}{\micro W}$ is applied, and the system is thermally equilibrated.
At the start of each individual measurement, the heating power is set to a reduced value following the heating protocol $P_\mathrm{MC} = 600,~300,~160,~80,~30,~10,~\SI{0}{\micro W}$, remaining constant during the measurement.
The electron temperature reduces with time, as the \gls{QD} system thermodynamically equilibrates after a change of the heating power.
Before any heating power is applied, the barriers are carefully adjusted at the lowest temperature, as described in section~\ref{sec:QD_tuning}, such that the \gls{QD} is weakly coupled to its leads. The resonances are thermally broadened, and the lever arm $\alpha_\mathrm{PG}$ is determined by a finite bias measurement [c.f. Fig.~\ref{fig1}(e)].

% explain data set at a fixed magnetic field
In Fig.~\ref{fig5}(b), we show the conductance as a function of plunger gate for five different times indicated in Fig.~\ref{fig5}(a).
In the integer quantum Hall regime, the electronic transport of a large \gls{QD} is expected to occur through a single level \cite{roosli_observation_2020}.
The conductance as a function of the plunger gate voltage $V_\mathrm{PG}$ follows \cite{averin_theory_1991,beenakker_theory_1991}
\begin{equation}
G_\mathrm{SL}(V_\mathrm{PG},T_\mathrm{e}) = G_0\cosh^{-2}\left[\dfrac{e \alpha_\mathrm{PG}(V_\mathrm{PG}-V_0)}{2k_\mathrm{B}T_\mathrm{e}}\right]\,,
\label{eq:single_level_temp}
\end{equation}
where $G_0 \propto 1/T_\mathrm{e}$ and $V_0$ describes the gate voltage where the resonance assumes its maximum value.
This dependence is valid if the coupling rate of the \gls{QD} to the leads, $\Gamma_\mathrm{tun}$, is weak compared to temperature $T_\mathrm{e}$, while the single-particle level spacing, $\Delta E$, dominates over temperature, i.e.  $h\Gamma_\mathrm{tun} \ll k_\mathrm{B} T_\mathrm{e} < \Delta E$.
The width of the peak $\Gamma_\mathrm{FWHM}$ (full width at half maximum) is directly related to the electronic temperature $T_\mathrm{e}$ by $\Gamma_\mathrm{FWHM} = 4\arcosh{(\sqrt{2})}k_\mathrm{B}T_\mathrm{e} \approx 3.53 k_\mathrm{B}T_\mathrm{e}$.
Neighbouring conductance resonances start to overlap for higher temperatures (or at higher magnetic fields). Therefore, we fit the resonances including the two next-neighbouring peaks
\begin{equation}
\begin{aligned}
	G_\mathrm{dot}(V_\mathrm{PG},T_\mathrm{e}) = & G_\mathrm{SL}(V_\mathrm{PG},T_\mathrm{e})
	+ G_\mathrm{SL}(V_\mathrm{PG}-\Delta V_\mathrm{PG},T_\mathrm{e}) \\
	& + G_\mathrm{SL}(V_\mathrm{PG}+\Delta V_\mathrm{PG},T_\mathrm{e})\,,
	\label{eq:multipeak_single_level}
\end{aligned}
\end{equation}
where the separation in plunger gate voltage between neighboring  resonances, $\Delta V_\mathrm{PG}$, is determined from finite bias measurements, and the same amplitude $G_0$ is assumed for all three peaks [c.f. Eq.~(\ref{eq:single_level_temp})].
At higher temperatures we assume a fixed experimental offset $G_\mathrm{off}$ that is determined from fits at the lower temperatures, in order to minimize effects of the overlapping tails.
By fitting Eq.~(\ref{eq:multipeak_single_level}) to all traces in Fig.~\ref{fig5}(a), we extract the temperature as a function of time, which is plotted in Fig.~\ref{fig5}(c). Fits to the selected traces are plotted in Fig.~\ref{fig5}(b) (solid lines).
For each measurement at fixed heating power, the temperature follows an exponential decay with time that is fitted (gray solid line) using $T_\mathrm{e}(t) = T_1 \exp(-\tau/t) + T_0$.
The exponential fits serve as a calibration of the electron temperature $T_\mathrm{e}$ as a function of time, and are used in later measurements employing the same heating protocol.

In addition, the dependence of the fitted peak amplitude $G_0$ on the electron temperature is shown in Fig.~\ref{fig5}(d).
The amplitude decreases with increasing temperature, following a $1/T_\mathrm{e}$ behavior which agrees with the expected single-level behavior \cite{averin_theory_1991, beenakker_theory_1991, roosli_observation_2020}.
This consistently matches the presented model, and supports the use of the temperatures extracted from Fig.~\ref{fig5}(a),~(c) as a reference for other measurements.
We did not observe qualitative changes in the behavior between resonances with different total electron number, when tuning the \gls{QD} as described in section~\ref{sec:QD_tuning}.

\begin{figure}[tbp]
\includegraphics[width=\columnwidth]{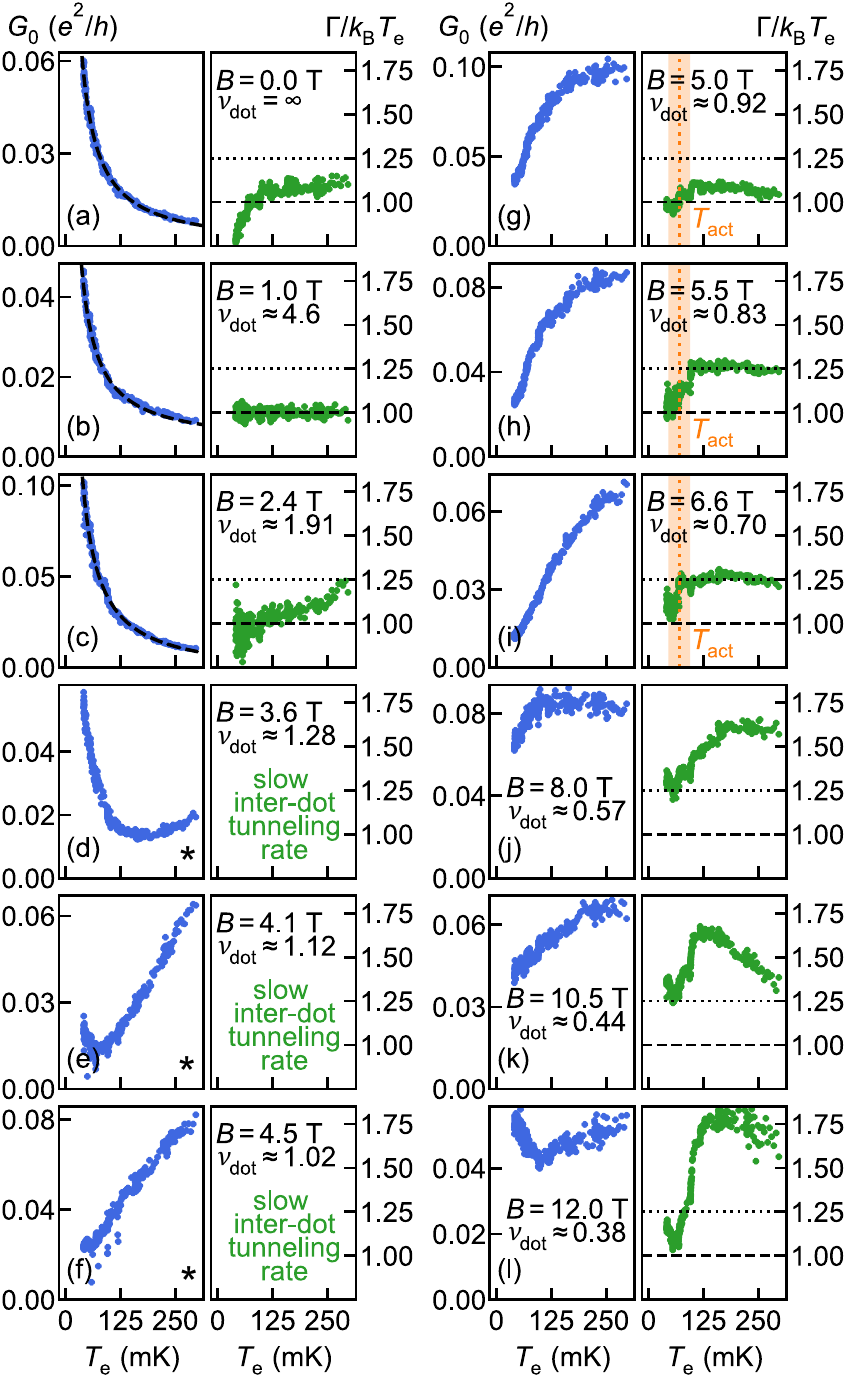}
\caption{Temperature dependence from $B = \SI{0}{T}$ to $\SI{12}{T}$. At each magnetic field value $B$, the left panel shows the peak amplitude $G_0$, and the right panel shows the normalized resonance width, $\Gamma/k_\mathrm{B}T_\mathrm{e}$, as a function of the electron temperature $T_\mathrm{e}$, respectively. Here, the full width at half maximum (FWHM) corresponds to $\Gamma_\mathrm{FWHM} = 4\arcosh{(\sqrt{2})}\Gamma \approx 3.53 \Gamma$.
	Applying the same measurement protocol as for the reference measurement at $B = \SI{1}{T}$ (see Fig.~\ref{fig5}), the corresponding electron temperature $T_\mathrm{e}$ as a function of time [Fig.~\ref{fig5}(c)] is used as a reference temperature at all magnetic field values.
	For (d)--(f) the interdot tunneling rate between the two compressible regions is low compared to the measurement timescale. Here, the peak amplitude is approximated with the maximum conductance of the trace, while the peak width cannot be determined (indicated by *).
	We fit the peak amplitude in (a)--(c) with a $1/T_\mathrm{e}$-dependence corresponding to single-level transport (black dashed line).
	For (d)--(i) the amplitude transitions continuously from a decreasing to an increasing dependence with increasing temperature.
	At (a)--(c) the normalized peak width $\Gamma/k_\mathrm{B}T_\mathrm{e}$ lies around 1 (dashed line), while stabilizing around 1.25 (dotted line) at higher temperatures for (h) and (i). In (g)--(i), the normalized peak width transitions from 1 towards 1.25 at the activation temperature $T_\mathrm{act} = \SI{70 \pm 25}{mK}$ (dashed orange line with uncertainty region shaded).}
\label{fig6}
\end{figure}

% temperature dependence for different magnetic fields
We continue by studying the evolution of the temperature dependence of the conductance resonances with magnetic field.
In Fig.~\ref{fig6}, we show data for the accessible magnetic field range $B = \SI{0}{T}$ to $\SI{12}{T}$, covering the integer and fractional quantum Hall regime.
Every subfigure in Fig.~\ref{fig6} corresponds to a dataset of the same form as the reference measurement displayed in Fig.~\ref{fig5}, at a different fixed magnetic field and filling factor $\nu_\mathrm{dot}$.
In the left panel, we plot the dependence of the fitted peak amplitude $G_0$ on the electron temperature $T_\mathrm{e}$, where the resonance is fitted according to Eqs.~(\ref{eq:single_level_temp}) and (\ref{eq:multipeak_single_level}).
The right panel shows the width of the fitted resonance normalized with the electron temperature, i.e.,  $\Gamma/k_\mathrm{B}T_\mathrm{e}$, as a function of the temperature, where the full width at half maximum (FWHM) is given by $\Gamma_\mathrm{FWHM} = 4\arcosh{(\sqrt{2})}\Gamma \approx 3.53 \Gamma$.
Applying the same heating protocol as in the reference measurement (see Fig.~\ref{fig5}), we use the dependence of the electron temperatures on time extracted in Fig.~\ref{fig5}(c), as a calibration at all magnetic fields.
At the magnetic fields shown in Fig.~\ref{fig6}(d)--(f) (marked with *), the interdot tunneling rate between the two compressible regions is slow compared to the measurement time scale \cite{van_der_vaart_time-resolved_1994, van_der_vaart_time-resolved_1997, roosli_observation_2020}.
In this case, fitting the resonances is not possible as the quasiparticles tunneling between the two compressible regions cause discrete jumps in the conductance.
Instead, we estimate the peak amplitude as the maximum conductance value of the trace.
Furthermore, the width of the resonance cannot be determined. 

% fitting formula multiple levels
When multiple levels contribute to electric transport, the conductance resonance shape follows \cite{beenakker_theory_1991}
\begin{equation}
\begin{aligned}
	G_\mathrm{ML} &= G_0 \dfrac{e \alpha_\mathrm{PG}(V_\mathrm{PG}-V_0)}{k_\mathrm{B}T_\mathrm{e}}\sinh^{-1}\left[\dfrac{e \alpha_\mathrm{PG}(V_\mathrm{PG}-V_0)}{k_\mathrm{B}T_\mathrm{e}}\right]\\
	&\approx G_0\cosh^{-2}\left[\dfrac{e \alpha_\mathrm{PG}(V_\mathrm{PG}-V_0)}{2.5k_\mathrm{B}T_\mathrm{e}}\right]\,,
\end{aligned}
\label{eq:multi_level_temp}
\end{equation}
which can be approximated within 1\% by the single level formula in Eq.~(\ref{eq:single_level_temp}) with a scaled peak width, i.e. temperature, by a factor 1.25.
This allows us to fit all resonances in Fig.~\ref{fig6} with the peak shape corresponding to single-level transport given in Eq.~(\ref{eq:single_level_temp}).
The ratio $\Gamma/k_\mathrm{B}T_\mathrm{e}$ of the resulting peak width $\Gamma = k_\mathrm{B}T_\mathrm{fit} =\Gamma_\mathrm{FWHM}/(4\arcosh{(\sqrt{2})})$ with the reference electron temperature $T_\mathrm{e}$, provides an indication to distinguish single-level transport, where $\Gamma/k_\mathrm{B}T_\mathrm{e} \approx 1$, and multi-level transport, where $\Gamma/k_\mathrm{B}T_\mathrm{e} \approx 1.25$.
In addition, the peak amplitude $G_0$ is mostly independent of temperature for multi-level transport, as the number of contributing level increases approximately with $T_\mathrm{e}$, and compensates for the $1/T_\mathrm{e}$-dependence of the amplitude of a single level \cite{beenakker_theory_1991}.

% integer filing factors to 2/3
For filling factors $\nu_\mathrm{dot} \gtrsim 2$ shown in Fig.~\ref{fig6}(a)--(c), the peak amplitude follows a $1/T_\mathrm{e}$-dependence (fit indicated by the black dashed line), which corresponds to single-level transport.
The normalized peak width $\Gamma/k_\mathrm{B}T_\mathrm{e}$ is around 1, where small deviations are observed for the lowest temperatures at $B = \SI{0}{T}$ [Fig.~\ref{fig6}(a)], which are slightly lower than the reference temperatures at finite magnetic field.
For $B = \SI{2.4}{T}$ [Fig.~\ref{fig6}(c)] the normalized peak width starts to increase at elevated temperatures.
At higher magnetic fields in Fig.~\ref{fig6}(d)--(e), we observe a smooth transition where the peak amplitude starts to increase at higher temperatures after initially decreasing at the lowest temperatures.
The temperature where the increase sets in, reduces with increasing magnetic field, until a linearly increasing behavior is reached in Fig.~\ref{fig6}(f) at  $B = \SI{4.5}{T}$ ($\nu_\mathrm{dot} \approx 1$).
This evolution coincides with the reduction of the charging energy $E_\mathrm{C}$ and the lever arm $\alpha_\mathrm{PG}$ by approximately a factor of two, which happens in the same magnetic field range, as discussed before [see Fig.~\ref{fig2}(b)].
The peak amplitude starts to saturate at higher temperatures, when further increasing the magnetic field to filling factors $\nu_\mathrm{dot} < 1$ in Fig.~\ref{fig6}(g)--(i).
For this magnetic field regime, the normalized peak width approaches 1.25 at higher temperatures, indicating multi-level transport, whereas for low temperatures the normalized resonance width decreases towards 1.
The transition, where the normalized resonance width starts to deviate clearly from 1, is associated with an activation temperature $T_\mathrm{act} =\SI{70\pm25}{mK}$ [orange dashed line in Fig.~\ref{fig6}(g)--(i)].
The observed evolution with increasing magnetic field is very similar to the temperature dependence of the amplitude of a \gls{QD} that transitions from single-level to multi-level transport, as calculated by Van Houten, Beenakker and Staring \cite{van_houten_coulomb-blockade_1992}. An initial $1/T_\mathrm{e}$-dependence (single-level transport) evolves into an increase, before saturating at high temperatures (multi-level transport). 

% Luttinger liquid model and interpretation
The unexpected temperature dependence of the conductance for $\nu_\mathrm{dot} \leq 1$ can be interpreted in terms of a Luttinger liquid model for edge states.
To apply such a model, we first assume that the excitation energies of the edge mode inside the dot are quantized with an energy spacing of $h v/L$, and that the temperature is much smaller than this spacing.
Then, the quantum dot can be modeled as a single resonant level for electrons passing through the dot, and the transmission probability through a dot level at energy $\epsilon_0$ can be described by a Breit-Wigner resonance $\Gamma_\mathrm{L} \Gamma_\mathrm{R} \over (\epsilon - \epsilon_0)^2 + \Gamma_\mathrm{T}^2$.
Here, $\Gamma_\mathrm{L}$ and $\Gamma_\mathrm{R}$ denote the partial widths of the resonant dot level due to tunneling into the left and right lead, while $\Gamma_\mathrm{T} = (\Gamma_\mathrm{L} + \Gamma_\mathrm{R})/2$ is the full width. When tunneling from an integer edge state described by a chiral Fermi liquid into the dot, the $\Gamma_{L/R}$ are independent of temperature, and Eq.~(\ref{eq:single_level_temp}) describes the conductance through the \gls{QD} with $G_0 \propto 1/T_\mathrm{e}$.
However, when tunneling from a fractional edge state, described by a chiral Luttinger liquid, onto the resonant level, the partial widths become temperature dependent according to  \cite{kane_edge-state_1996}
%
%***************************  temperature dependence of partial width  **********
\begin{equation}
\Gamma_\mathrm{L/R} = \Gamma_\mathrm{0,L/R} \left({T_\mathrm{e}\over T_0}\right)^{m-1} \,,
\label{partialwidth.eq}
\end{equation}
%************************************************************************************
%
where $\Delta \nu = 1/m$ is the filling fraction difference characterizing the edge state, and $T_0$ is a high energy cutoff.
For example, the edge mode between the  vacuum and and an incompressible Laughlin state at filling 1/3 would have $\Delta \nu = 1/3$ and $m=3$. Since the prefactor in Eq.~(\ref{eq:single_level_temp}) is $G_0 \propto \Gamma_\mathrm{L} \Gamma_\mathrm{R}/(\Gamma_\mathrm{T} k_\mathrm{B} T_\mathrm{e})$, the temperature dependence becomes $G_0 \propto T_\mathrm{e}^{m-2}$.
For the above example of a $1/3$ edge channel with $m=3$ one thus finds $\Gamma_0 \propto T_\mathrm{e}$. 

From the experimental data in Fig.~\ref{fig1}(c) one sees that the bulk is on the $\nu_\mathrm{bulk}=1$ integer plateau when the dot filling is in the range $0.7 \leq \nu_\mathrm{dot} \leq 1$.
In the following, we assume that the integer edge undergoes reconstruction with an outer $1/3$-channel, at least in the vicinity of the \glspl{QPC}, where the electron density is reduced with respect to the the bulk value.
Experimental evidence for such a reconstruction has been reported in Ref.~\cite{roddaro_particle-hole_2005}, where $\Delta \nu = 1/3$ renormalization for tunneling through a \gls{QPC} was found for a bulk filling $\nu_\mathrm{bulk} = 1$.
In such a scenario, tunneling into the dot takes place from an outer $1/3$-channel, and we can use $m=3$ in Eq.~(\ref{partialwidth.eq}) to obtain $G_0(T_\mathrm{e}) \propto T_\mathrm{e}$, which is in good agreement with the data presented in Fig.~\ref{fig6}(i).

%COMMENT On the other hand, for a dot filling factor $\nu_{dot} \approx 0.5$, the bulk assumes an incompressible  $\nu = 2/3$ state [see Fig.~\ref{fig1}(c)]. Theory \cite{kane_randomness_1994,kane_edge-state_1996,shytov_tunneling_1998} predicts an exponent $m=2$ for tunneling into a well equilibrated $\nu = 2/3$ edge. Using this choice in Eq.~(\ref{eq:single_level_temp}), we find that $G_0(T_\mathrm{e})$ should be temperature independent, in agreement with Fig.~\ref{fig6}(j). Earlier experiments \cite{grayson_continuum_1998,chang_plateau_2001} observed smaller values of the Luttinger liquid exponent when studying tunneling from a 2DEG in to a sharp edge of bulk filling factor $2/3$, obtained via cleaved-edge-overgrowth.

From the temperature dependence of the level width in Fig.~\ref{fig6}(i), we can estimate the edge velocity of a fractional $2/3$ mode: at an activation temperature $T_\mathrm{act}\approx \SI{70}{mK}$ [see orange dotted line in Fig.~\ref{fig6}(g)--(i)], a crossover from a normalized resonance width $\Gamma/k_\mathrm{B} T_\mathrm{e} \approx 1$ to a peak width $\Gamma/k_\mathrm{B} T_\mathrm{e} \approx 1.25$ is observed, indicative of a crossover between tunneling into a single level or multiple levels, respectively. 
When describing the  edge  mode inside the quantum dot as a Luttinger liquid, then the multi-level regime corresponds to the excitation of plasmons with an excitation energy of $h v_\mathrm{edge}/L$.
Here, $L\approx\SI{3.05}{\micro m}$ \cite{roosli_fractional_2021} denotes the radius of the edge mode for $\nu_\mathrm{dot} = 2/3$.
Relating excitation energy and crossover temperature via $h v_\mathrm{edge}/L \simeq 3.53  k_\mathrm{B} T_\mathrm{act}$, we obtain the estimate $v_\mathrm{edge} = \SI{1.6e4}{m/s}$. This is somewhat smaller than the recently reported edge velocities for plasmons of $5$--$\SI{8e4}{m/s}$ for a 2/3 edge and $2$--$\SI{4e4}{m/s}$ for a 1/3 edge, obtained in  time-of-flight measurements of a charge wave package \cite{lin_quantized_2021}.
Interferometry experiments determined the edge velocity of integer edge modes to $0.5-\SI{2e5}{m/s}$ \cite{mcclure_edge-state_2009, nakamura_aharonovbohm_2019}, which is larger than our value for fractional edge modes.
However, it is not unexpected that the velocity of fractional modes is smaller than that of integer modes: 
first, the edge velocity in the presence of Coulomb interactions is expected to be proportional to the filling fraction difference $\Delta \nu$ of the mode, leading to a moderate reduction of the $2/3$ velocity as compared to the integer velocity.
Second, due to the smaller energy gap of the $2/3$ state as compared to an integer state, the formation of localized quasi-particles is energetically easier in the fractional state than in the integer state.
In the presence of a tunnel coupling of such localized states to the ideal edge mode, a particle travelling along the edge will spend some time on the localized states, which will further reduce the velocity of the edge mode.
Combining above arguments, the velocity determined for the $2/3$ edge mode is plausible compared with the previously observed integer and fractional edge velocities \cite{lin_quantized_2021, mcclure_edge-state_2009, nakamura_aharonovbohm_2019}. 

In theory, the $2/3$ edge is predicted to develop one \cite{kane_randomness_1994} or several \cite{wang_edge_2013} neutral modes in the presence of disorder and Coulomb interaction, where neutral modes can have a reduced velocity compared to that of the charge mode.
Experimental evidence for such a neutral mode was found in the form of upstream heat transport \cite{bid_observation_2010}. 
If the excitation energy of the neutral mode was larger than the thermal broadening of peaks, a bunching of transmission resonances would result due to the fact that the electron operator excites the neutral mode \cite{kane_randomness_1994}.
We have not observed such bunching in the regime of $\nu_\mathrm{dot} \approx 2/3$. Therefore, we conclude that the energy scale of neutral modes is smaller than the peak broadening even at the lowest temperatures studied, i.e., $\SI{50}{mK}$. 

% unreliable beyond 8T
For magnetic fields $B > \SI{8}{T}$ in Fig.~\ref{fig6}(j)--(l), we cannot identify clear trends for the amplitude of the resonances, while the width varies strongly.
At this point, the gate voltage of the barrier gates need to be adapted, in order to main the desired weak coupling of the \gls{QD} to the leads [c.f. Fig.~\ref{fig2}(a)].
We think that the tunneling through the barriers is dominated by localized states as discussed above.
In addition, the tuning of the barriers strongly depends on temperature in this regime.
We assume that localized states close to the barriers are thermally activated, resulting in increased coupling at higher temperatures.
The strong coupling resonances are broadened beyond thermal broadening, leading to an increase of the normalized peak width $\Gamma/k_\mathrm{B}T_\mathrm{e} > 1.25$ at elevated temperatures.
On the other hand, resonances remain thermally broadened at low temperatures exhibiting a normalized width $\Gamma/k_\mathrm{B}T_\mathrm{e} \lesssim 1.25$ for the same barrier gate voltages.
In this magnetic field regime, the temperature dependence of the conductance is dominated by the tunneling barriers.

To summarize, we studied the varying temperature dependence of the Coulomb resonances of a \gls{QD} in the integer and fractional quantum Hall regime, which is consistent with a Luttinger liquid model interpretation for $\nu_\mathrm{dot} < 1$.

\section{Bouncing states at B = 0~T}
\label{sec:bouncing_states}

\begin{figure}[tbp]
\includegraphics[width=\columnwidth]{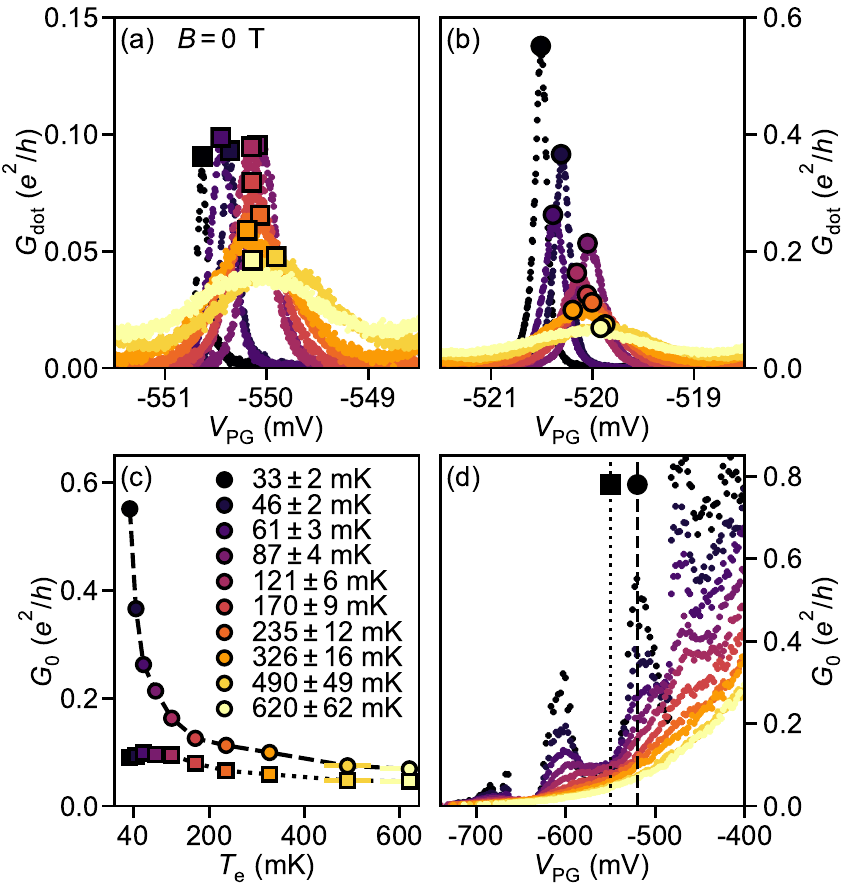}
\caption{Temperature dependence at magnetic field $B = \SI{0}{T}$: (a) and (b) show the conductance $G_\mathrm{dot}$ as a function of the plunger gate voltage at different electronic temperatures $T_\mathrm{e}$ for two Coulomb peaks, centered around two different values of plunger gate voltage. The maximal conductance is plotted in (c) against electron temperature $T_\mathrm{e}$ for the two resonances in (a) and (b). (d) shows the maximal conductance of approximately 130 resonances at different plunger gate voltages for various temperatures, where each point corresponds to a single resonance at a certain temperature. While some resonances show a quickly decreasing amplitude with increasing temperature [as for (b), circles in (c)], others show a small increase followed by a weak decrease [as for (a), squares in (c)]. The temperature is varied by applying constant heating power to the mixing chamber. The electron temperature $T_\mathrm{e}$ is determined by fitting the width of Coulomb resonances in a reference measurement at $B = \SI{0}{T}$ (the highest two temperature values are deduced from a ruthenium oxide resistor at the mixing chamber). } 
\label{fig7}
\end{figure}

For completeness, we present the temperature dependence at zero magnetic field, where the amplitude of individual Coulomb resonances exhibit different behaviour with temperature.
Figure~\ref{fig7}(a) and (b) shows the conductance of two  separate resonances centered around different values of the plunger gate voltage for various electron temperatures $T_\mathrm{e}$.
The electron temperature is changed by applying a constant heating power to the mixing chamber, and is determined from the width of the conductance resonances in a reference measurement.
The resonance amplitude changes with temperature, and the peak positions vary slightly between individual measurements at different temperatures, which we associate with changes due to heating and discrete charge fluctuations over time.
At the edges of the presented range, the tails of the next-neighboring peaks overlap for elevated temperatures, leading to an increased conductance in the Coulomb blockade region.
The conductance maxima of the resonances at a certain temperature are indicated [Fig.~\ref{fig7}, squares (a) or circles (b)], and plotted in Fig.~\ref{fig7}(c) as a function of the electron temperature.
The two resonances show distinctly different evolution of the conductance amplitude with increasing temperature:
The resonance in Fig.~\ref{fig7}(b) shows a strong decrease with temperature saturating towards higher temperatures,
while the resonance in Fig.~\ref{fig7}(a) is slightly increasing at low temperatures, and only weakly decreases at higher temperatures.
The behavior of the resonance in Fig.~\ref{fig7}(b) corresponds qualitatively to the single-level transport behavior previously discussed for Fig.~\ref{fig6}(a)--(c).

Fig.~\ref{fig7}(d) shows the  conductance maximum of roughly 130 resonances in a wide range of the plunger gate voltage at different electron temperatures, where each measurement point corresponds to the maximum conductance value of one resonance at a certain temperature.
The dotted (a) and dashed (b) line mark the two resonances presented in Fig.~\ref{fig7}(a)--(c), respectively.
The conductance maxima are modulated by an envelope function that shows four broad maxima at the lowest temperatures, while it is monotonically increasing with plunger gate voltage at higher temperatures. The barrier gate voltages are fixed during the measurement.
Therefore, the tunnel coupling across the barriers increases with plunger gate voltage, resulting in a general increase of the conductance at fixed temperature for higher plunger gate voltages. 
At the maxima of the envelope function, the resonance amplitudes are strongly decreasing with temperature, as observed for the resonance in Fig.~\ref{fig7}(b). In contrast, at the minima of the envelope, the resonance amplitudes are less dependent on temperature, exhibiting behavior similar to the resonance in Fig.~\ref{fig7}(a).

Previous experiments reported that the amplitude and shape of Coulomb resonances can be influenced by \gls{QD} states that couple strongly to the leads, so called `bouncing states'\cite{lindemann_bouncing_2002-1,hackenbroich_bouncing-ball_2000}. Here, we relate the modulation of the peak amplitude to bouncing states existing at the centers of the maxima of the envelope of the conductance. % In addition, we observe  a slightly asymmetric peak shape that inverts when crossing a maximum in the envelope, such that the steep slope of the peak faces towards the maximum. 
We assume that the presence of bouncing states influences the temperature dependence of the resonance amplitudes at $B = \SI{0}{T}$. Thereby, resonances close to a bouncing state show a strongly decreasing, 1/$T_\mathrm{e}$-like dependence with temperature, while resonances away from a bouncing state show a constant or slowly decreasing behavior.

\section{Conclusion}
\label{sec:conclusion}
% Conclusion
In this paper, we have studied in detail a weakly coupled, large \gls{QD} in different magnetic field regimes.
In the integer and fractional quantum Hall regime, the conductance resonances are modulated by internal charge rearrangements, resulting from the charge reconstruction into concentric compressible and incompressible regions inside the \gls{QD}. While previously measured for $2 > \nu_\mathrm{dot} > 1$ \cite{van_der_vaart_time-resolved_1994,van_der_vaart_time-resolved_1997,heinzel_periodic_1994, fuhrer_transport_2001, chen_transport_2009-1, baer_cyclic_2013, sivan_observation_2016-1, liu_electrochemical_2018, roosli_observation_2020} and $\nu_\mathrm{dot} \gtrsim 2/3$, $1/3$ \cite{roosli_fractional_2021}, we observe complex patterns of modulated resonances for higher integer filling factors $6 > \nu_\mathrm{dot} > 2$, where quasiparticles tunnel between more than two compressible regions.
We extract the magnetic spacings between resonances, and find reduced periodicities, pointing to cyclic depopulation of the innermost compressible region with magnetic field.
In addition, the signal of an external electrostatic charge detector allows to measure the internal charge rearrangements in the \gls{QD} for the integer quantum Hall regime. While the charge detection signal agrees with the direct \gls{QD} conductance, we do not observe additional charging lines. Charge sensing was not achieved in the fractional quantum Hall regime due to reduced tunability and missing sensitivity of the charge detector at high magnetic fields, which might be resolved using an adapted sample design.
Furthermore, we studied the temperature dependence of the conductance resonances for magnetic fields ranging from \SI{0}{T} to \SI{12}{T}.
At low magnetic fields and in the integer quantum Hall regime, we find a $1/T_\mathrm{e}$-dependence of the conductance resonance amplitude with temperature, corresponding to single-level transport.
For a filling factor, $\nu_\mathrm{dot}$, between 2 and 1, this dependence continuously transitions into a linear dependence of the amplitude on temperature.
This coincides with a reduction of the charging energy (and the lever arm) by a factor of approximately 2 in the same regime, while the plunger gate capacitance roughly stays constant.
For fractional filling factors $\nu_\mathrm{dot} < 1$, the measured temperature dependence of the conductance resonances is consistent with an interpretation using a Luttinger liquid model.
At $B = \SI{0}{T}$, the temperature dependence of individual resonances varies, as it is influenced by `bouncing states' \cite{lindemann_bouncing_2002-1}, which are coupling strongly to the leads.
Our observations extend and complement the picture of internal quasiparticle tunneling in a \gls{QD} in the quantum Hall regime.

\begin{acknowledgments}
The authors acknowledge the support of the ETH FIRST laboratory and the financial support of the Swiss Science Foundation (Schweizerischer Nationalfonds, NCCR QSIT). B.~R. would like to acknowledge support by DFG grant RO 2247/11-1. This manuscript is included as a part in the doctoral thesis of M.~R. at ETH Zurich, freely accessible at \href{https://doi.org/10.3929/ethz-b-000511370}{doi:10.3929/ethz-b-000511370}.
\end{acknowledgments}

% bibliography, for uploading direct inclusion of .bbl file is preferred
%\bibliography{bibl}

\end{document}